\documentclass[12pt,preprint]{aastex}

\RequirePackage{latexsym}%
\RequirePackage{graphicx}%
\RequirePackage{amssymb}%
\RequirePackage{natbib}%

\newcommand{\jybm}{\mbox{Jy~beam${}^{-1}$}}

\newcommand{\htwo}{\ion{H}{2}}
\newcommand{\cpap}{\clearpage}
\newcommand{\uv}{$u$-$v$}

\newcommand{\etal}{et al.}
\newcommand{\kpc}{Kelvin~pc$^{-1}$}
\newcommand{\ep}{\varepsilon}

\shorttitle{HII Regions in Absorption at 74 MHz}
\shortauthors{Nord et al.}

\begin{document}

\title{74 MHz Discrete \htwo\ Absorption Regions Towards The Inner Galaxy}

\author{Michael E. Nord}
\affil{Naval Research Laboratory, Remote Sensing Division, Code 7263, 4555 Overlook Avenue SW,  Washington, DC
	20375-5351}
\email{Michael.Nord@nrl.navy.mil}

\author{P.A. Henning, R.J. Rand}
\affil{Department of Physics and Astronomy, University of New Mexico,
	800 Yale Boulevard NE, Albuquerque, NM  87131}
\email{henning@phys.unm.edu}
\email{rjr@phys.unm.edu}

\and

\author{T.~Joseph~W.~Lazio, Namir~E.~Kassim}
\affil{Naval Research Laboratory, Remote Sensing Division, Code 7213, 4555 Overlook Avenue SW,  Washington, DC
	20375-5351}
\email{Joseph.Lazio@nrl.navy.mil}
\email{Namir.Kassim@nrl.navy.mil}

\begin{abstract}

At low radio frequencies ($\nu \lesssim 100$ MHz), classical \htwo\ regions
may become optically thick (optical depth $\tau \ge 1$) and can be observed as discrete absorption regions
against the Galactic nonthermal background emission created by Galactic
cosmic ray electrons spiraling around magnetic fields.  However, the historically poor angular resolution ($>30\arcmin$) of previous low frequency surveys has limited such observations to the largest and nearest \htwo\ regions.  The significantly enhanced resolution and surface brightness sensitivity of the 74 MHz system on the VLA now allow for the detection of absorption regions on scale sizes of just a few arcminutes that can be readily identified with many more \htwo\ regions previously cataloged in emission at higher frequencies.  These absorption measurements directly constrain the brightness temperature of the cosmic ray synchrotron emission emanating from behind the \htwo\ regions based on reasonable physical assumptions.  Many such observations could be used to map out the
3-dimensional cosmic ray emissivity in the Galaxy
without resorting to \textit{a priori} assumptions about Galactic structure.
This measurement is unique to low frequency radio astronomy.

In this work we present 74 MHz observations in the region
$26\arcdeg>l>-15\arcdeg$, $-5\arcdeg<b<5\arcdeg$, report the detection of 92
absorption features associated with known \htwo\ regions, and derive the brightness
temperature of the Galactic cosmic ray electron synchrotron emission emanating from the column behind these
regions.  For the 42 \htwo\ regions with known distances, the average
emissivity of the column behind the \htwo\ region is derived.  74 MHz emissivity
values range between 0.3 and 1.0 \kpc\ for a model assuming uniform distribution of emissivity.  
Methods for utilizing this type of data to model the 3-dimensional distribution of cosmic ray emissivity and the possibility of using this method to break the \htwo\
region kinematic distance degeneracy are discussed.

\end{abstract}


\section{Introduction}

At low frequencies, the Galactic background emissivity\footnote{Emissivity is a measure of
intensity per unit length (SI units are W M$^{-3}$ Hz$^{-1}$ ster$^{-1}$).  As
equivalent brightness temperature will be used in this work, the units used will be
Kelvin~pc$^{-1}$ and the symbol $\ep$ used to denote it.} is dominated by the nonthermal synchrotron
radiation caused by cosmic ray (CR) electrons interacting with the Galactic
magnetic field.  Though CR's were first observed nearly 90 years ago
(Hess 1919) and are thought to represent a significant fraction of the energy
density of interstellar space ($\sim1.8$ eV~cm$^{-3}$, Webber 1998), their origin, distribution, and energy spectrum are still only partly understood.  This is in part because though multiple tracers of CR electrons exist, they typically produce
line-of-sight integrated measurements only.  One must then
assume a model of Galactic structure to derive the three-dimensional distribution of the CR emissivity.

It is now widely accepted that Galactic supernova remnants (SNRs) are responsible for the acceleration of CRs of energies up to at least $10^{15}$ eV (e.g.~Jones et al.~1998), a view supported by the detection of nonthermal X- (Koyama et al. 1995) and $\gamma$-ray (Aharonian et al. 2005) emission from SNRs.  The typical energy of the CR electron population being detected at 74 MHz may be calculated by assuming the CR energy spectrum is a power law of the form $N(E)\propto E^{-2.5}$ (Webber~1993a).  Relativistic electrons with energy $E$ emit maximum power at a frequency of
$$\nu = 16(MHz) B(\mu G) E^2(GeV)$$ (Rockstroh \& Webber~1978),  resulting
in a typical energy of $\sim$0.4 GeV at 74~MHz for a reasonably assumed mean Galactic magnetic
field strength of $5\mu$G (e.g. Rand and Kulkarni 1989).  Therefore SNRs are the mostly likely progenitor of many if not most of the CR electrons in the energy range we are sensitive to in this work.

After initial acceleration in SNR shock fronts, propagation of CRs may occur
through diffusion, convection, or through other mechanisms (e.g.~Duric 1999; Lisenfeld \& V\"{o}lk 2000; Webber 1990; Longair 1990; Cesarsky~1980), but a central problem in
placing observational constraints on all CR acceleration and propagation models  is the lack of distance
information available in CR tracer observations.

Observations of \htwo\ regions in absorption against the Galactic nonthermal
background date back to at least the early 1950's (Scheuer \& Ryle 1953), and many surveys of the low frequency radio sky have detected \htwo\ absorption (somewhat more recent examples include
Kassim 1988; Dwarakanath \& Udaya Shankar 1990; Braude, Sokolov, \& Zakharenko
1994).  However, the angular resolution of these surveys has always been poor ($>
30\arcmin$) due to the limitations of past low frequency telescopes.  Observations with the 74 MHz receiver system on NRAO's VLA
telescope (e.g.~Kassim et al.~1993) now permit much higher-resolution
imaging.  The $\sim10\arcmin$ beam used here achieved a compromise between spatial resolution and surface brightness sensitivity to detect many more discrete \htwo\ regions than previously possible.  Sub-arcminute resolution\footnote{10\arcsec\ resolution is possible in the A configuration + Pietown link mode e.g.~Gizani et al.~2005} is possible with the system, however the surface brightness sensitivity is insufficient to detect \htwo\ regions on size scales much larger than a few arcminutes.  Nevertheless, even near arcminute resolution is sufficient to establish a direct correspondence between discrete 74 MHz absorption features and many discrete \htwo\ regions identified in higher frequency emission surveys.

As will be shown in Section~\ref{sec:holes_theory} the high opacity of \htwo\
regions at low frequencies causes them to act as an opaque "wall" to
incident Galactic background radiation, allowing for the discrimination of radiation originating
behind and in front of the \htwo\ region.  Many such observations along many
lines of sight could allow a three dimensional CR emissivity map to be
constructed.  Such a map could be of great use in testing CR electron
production, diffusion, and aging models (e.g.~Longair 1990).  Observations at a
second frequency would allow for three-dimensional mapping of the spectrum of
CR emissivity.

This work is the first step in producing a comprehensive survey of discrete \htwo\ absorption with the aim of extracting information about the 3-dimensional distribution of the Galactic CR synchrotron emissivity.

\section{Observations and Analysis}\label{sec:obs}

Observations presented here were performed with the NRL/NRAO
74/330 MHz system on the VLA in the most compact configurations, which permit the optimal balance between surface brightness sensitivity and resolution for detecting the greatest number of \htwo\ absorption features.  Four fields were observed (field FWHM
$\sim 11\arcdeg$ at 74 MHz), centered on Galactic $(l,b)$ coordinates of
$(21.5, -0.9)$, $(11.2, -0.3)$, $(0,0)$, and $(349.7, +0.2)$.  The Galactic
Center region $(l,b=0,0)$ was observed by the VLA in D configuration on 31
May, 1999.  The duration of the
observation was about 7 hours, with approximately 5.5 hours of on-source time.
74~MHz and 330~MHz data were taken simultaneously.  The 74 MHz bandwidth was
1.5 MHz, which was split into 64 channels in order to facilitate RFI removal
and to reduce the effects of bandwidth smearing\footnote{After binning, the final frequency channel width was 93.75 kHz.  With an $\sim$8\arcmin\ synthesized beam, bandwidth smearing is $\sim$1\% effect at the edge of the primary beam}.  Three other pointings were observed cyclically in the DnC hybrid configuration 
on 28-29 September 2001.  The observation was 8 hours in
length.  However, the first half of the run was lost to especially poor ionospheric weather conditions.  Hence, approximately 45 minutes to an hour of on-source time per field
were obtained.

Several of the above fields were previously observed by Dr.~C.~Brogan at 74 MHz with
similar compact array configurations.  Since there is some \uv\ overlap in different
configurations, these data were included.  Table~\ref{tab:74observe} details
the observational parameters.

Maps of the Galactic center region at 330 MHz were presented in Nord et al.~(2004) and Nord (2005).  
Data reduction in this work is similar with a few notable exceptions.  
Amplitude calibration and bandpass calibration were performed using Cygnus~A.  Due to the paucity of strong point sources at 74~MHz and the large primary beam, standard VLA phase calibration is not practical at 74 MHz.  Instead, initial phase calibration was done using Cygnus~A, these phases were transferred to the field, and then residual phase changes due to ionospheric fluctuations were compensated for with self-calibration.  However, this can result in an overall phase shift and hence a positional shift on the sky.  For this reason, the first pass of phase self-calibration used a 330 MHz model provided by the data taken simultaneously with the 74 MHz data.  As ionospheric refraction effects scale as $\nu^{-2}$ (Erickson 1984), the 330 MHz model will suffer much less positional shift and result in much greater astrometric accuracy for the final image.  Because only nonthermal sources will be observed in emission at both frequencies, only 330~MHz sources that were known nonthermal emitters were included in the self-calibration model.

Radio Frequency Interference excision is a key issue in low frequency VLA observations.  Though algorithms exist that automate the removal of RFI, manual excision is the preferred method as excision algorithms can miss low level RFI.  First, visibilities with excessive amplitudes (e.g., $> 100\sigma$) were flagged.  Then the visibility data amplitudes were scrutinized in both Stokes I and V.  Stokes V is particularly useful in locating RFI as there should be very little circular polarization at these frequencies, while RFI is often highly circularly polarized.  An additional means by which RFI was localized was the identification of ripples in the image. Determining the spatial frequency of these ripples allowed the offending baseline and time range to be located and removed from the visibility ($u$-$v$) dataset.  See Nord et al.~(2004) and Lane et al.~(2005, in press) for a more extensive discussion on RFI excision.

After calibration and RFI excision, the data were averaged down to 14 channels.  This trade-off slightly increases bandwidth smearing, while significantly reducing the computational cost of imaging.  Faceted imaging (Cornwell \& Perley~1992) was then employed in order to mitigate non-coplanar effects in the image which are exacerbated by the large fields of view at low frequencies.  Approximately 40 facets per field were imaged, and a standard imaging / self calibration loop was used.  Each data set was reduced separately before the data from each field were combined.  One final amplitude and phase self-calibration was performed with a long solution interval in order to resolve any amplitude or phase differences between datasets.

As the fields were observed with differing VLA configurations, have differing $u$-$v$ coverage, and are located at different declinations, the size and shape of the synthesized beam changed between fields.  Because roughly equal surface brightness sensitivity is desired in order to compare the different fields, the final images were created utilizing a tapering function in the $u$-$v$ plane.  The $u$-$v$ data were weighted by a Gaussian having a $1/e$ point at a \uv\ separation of $700 \lambda$ ($\sim$2.8 km).  The resulting fields have similar resolutions, comparable to a C configuration observation, but have the surface brightness sensitivity of a D configuration observation.

\begin{figure}[htb]
\begin{center}
\includegraphics[angle=-90,width=\textwidth]{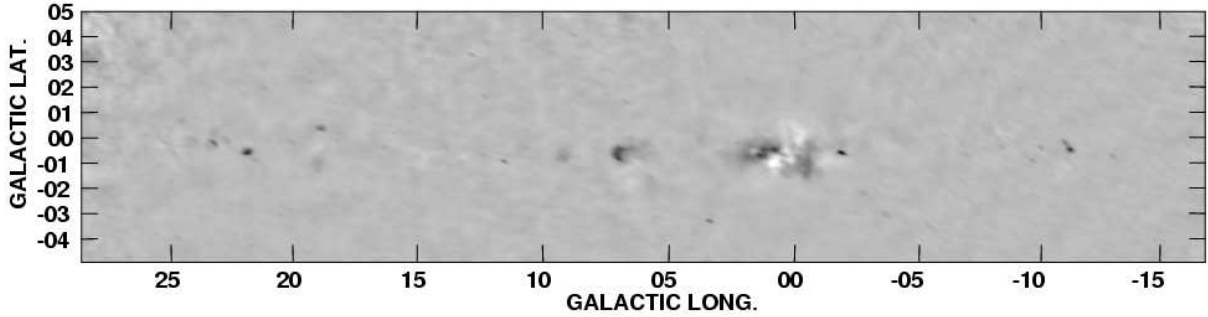}
\end{center}
\caption[74 MHz Galactic plane]{74 MHz image of the inner Galactic plane.  The image has a FWHM resolution of 14.8\arcmin\ by 6.2\arcmin\ with a position angle of 8\arcdeg.  The grey scale is linear from -40 to 110 Jy~beam$^{-1}$.  Dark regions indicate emission, white areas are regions of \htwo\ region absorption.  The image is meant only to orient the reader; measurements were made using data from individual pointings.  Due to smoothing, most of the \htwo\ regions detected in absorption in this work are not evident in this figure.}
\label{fig:holesmap}
\end{figure}

As the observed fields are spaced approximately one primary beam width from each other, one possible approach for comparing the fields would be to combine all the fields into one image.  This approach would have the advantage of greater sensitivity in the regions of overlap.  However, as all individual facets are convolved to have the same resolution before being combined, combining all four observed fields results in the final resolution being set by the lowest resolution facet.  This is unacceptable as the increased resolution and surface brightness sensitivity compared to previous low frequency observations is the main factor permitting the identification of many more individual \htwo\ regions.  For this reason, all measurements were done on the individual fields.  However, for the purpose of examining the entire field imaged, a low resolution image was created by combining all four fields.  The resulting image (Figure~\ref{fig:holesmap}) has a FWHM resolution of 14.8\arcmin\ by 6.2\arcmin\ with a position angle of 8\arcdeg\ and covers a Galactic latitude range of $\sim +5\arcdeg > b > -5\arcdeg$ and a Galactic longitude range of $\sim 26\arcdeg > l > -15\arcdeg$.

\subsection{Image Sensitivity}

The final rms image sensitivity of a 74 MHz VLA observation is not a simple function of observing time as it is at higher frequencies.  Though not as of yet fully quantified, the classical confusion limit of a VLA D configuration observation is estimated to be $\approx0.4$~\jybm (A. Cohen, private communication).  Furthermore, RFI and incomplete understanding of ionospheric effects can introduce distorted data, resulting in a situation where more data may actually adversely affect image quality.  Good \uv\ coverage and a calm ionosphere and RFI environment are more important than observation length.  The images created in the course of this work are approaching as low noise as the present VLA 74 MHz system will allow while maintaining the surface brightness sensitivity required to detect diffuse \htwo\ regions.  Addition of B and A configuration data could well decrease the confusion limit, but higher order ionospheric effects become very difficult to overcome in these configurations.  Further work in the detection and quantification of \htwo\ regions in absorption at low frequencies will likely have to await the high surface brightness sensitivity, high resolution, fully ionospherically compensated images expected out of the next generation of low frequency radio telescopes such as the Long Wavelength Array (LWA) and the Low Frequency Array (LOFAR).

\section{\htwo\ Absorption Region Interpretation}\label{sec:holes_theory}

Following Kassim (1987), a single dish radio telescope observing an \htwo\ region with an angular extent larger than the primary beam of the telescope would observe brightness temperature (degrees Kelvin): $$T_{h}=T_{e}(1-e^{-\tau})+T_{gb}e^{-\tau}+T_{gf}$$ where $T_{e}$ is the electron temperature of the \htwo\ region, $T_{gb}$ is the brightness temperature of the CR electron nonthermal emission behind the \htwo\ region, $T_{gf}$ is the brightness temperature of the CR electron nonthermal emission between the \htwo\ region and the observer, and $\tau$ is the optical depth of the \htwo\ region (any extragalactic contribution is assumed to be negligible).  For a typical Galactic \htwo\ region\footnote{$\tau(\nu) \sim 8.2\times10^{-2}T_e^{-1.35}\nu^{-2.1}$ times the emission measure (EM) for $T_e$ in Kelvin, $\nu$ in GHz, and EM in pc~cm$^{-6}$ (Harwit 1998).  For a typical $T_e\sim7000$K and EM$\sim10^5$ pc~cm$^{-6}$, $\tau\sim12.5$ at $\nu=$74 MHz.}, $\tau_{\mathrm{74 MHz}} \gg 1$, reducing the above equation to: $$T_{h}=T_{e}+T_{gf}$$  The single dish measuring a line-of-sight near the \htwo\ region would measure: $$T_{gt}=T_{gf}+T_{gb}$$ where $T_{gt}$ is brightness temperature from the Galactic total CR electron synchrotron emission.  As an interferometer is insensitive to the total flux, it measures the difference between these two, giving an observed brightness temperature on the \htwo\ region of: $$T_{obs}=(T_{e}+T_{gf})-(T_{gf}+T_{gb})  = T_{e}-T_{gb}$$  For typical electron and Galactic background temperatures ($T_e\approx7000$~K and $T_{gb}\approx20000$~K near the Galactic center) , we expect the observed $T_{obs}$ to be negative.

If the electron temperature of the \htwo\ region is either known from higher frequency measurements or can be estimated, the brightness temperature of the column \textit{behind} the \htwo\ region is then directly calculated.  If an appropriate measure of the total emission ($T_{gt}$) exists, the brightness temperature of the column between the \htwo\ region and the observer may be derived.  When the path length ($D$) over which the brightness temperature is integrated is known or can be deduced from reasonable physical assumptions, the emissivity may then be calculated ($\ep = T/D$ \kpc).   This ability to split the total line-of-sight emission
into columns behind and in front of the \htwo\ region allows information concerning the \textit{three-dimensional} distribution of Galactic CR emissivity to be extracted.  This technique is unique to the low frequency radio regime.

\section{A Catalog of HII Regions Detected in Absorption}\label{sec:holesdetect}

\begin{figure}[htb]
\begin{center}
\includegraphics[scale=.5,width=\textwidth]{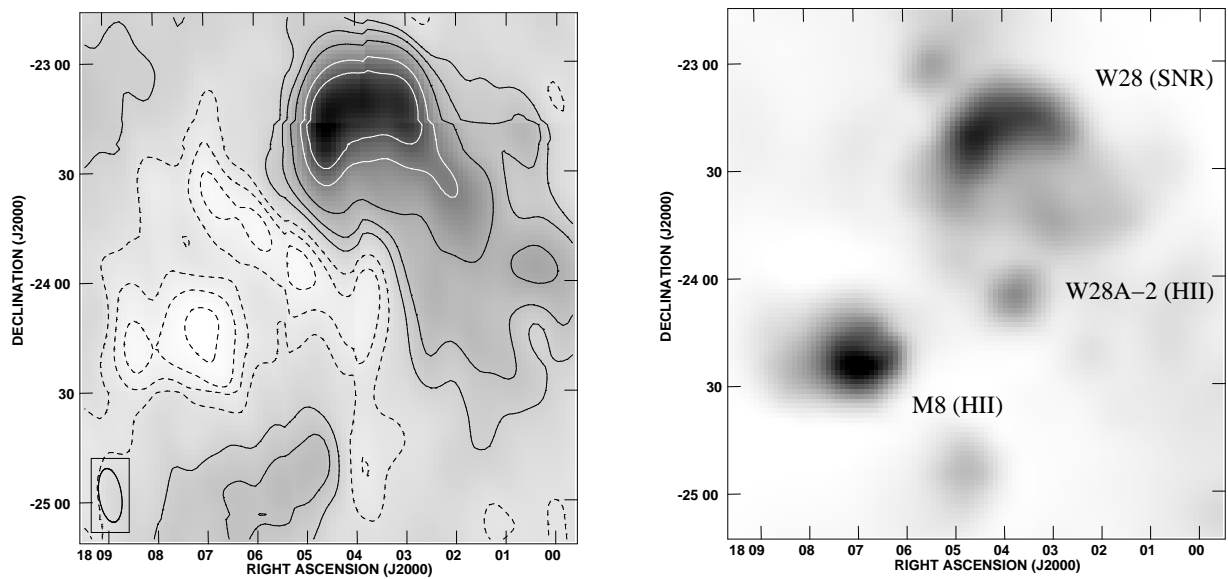}
\end{center}
\caption[74 MHz Image of the M8 Region]{\textit{Left:}  The region around the \htwo\ region M8 at 74 MHz.  Contours are -8, -4, -2, 2, 5, 10, 20 and 30 times the local rms of 1.4 \jybm (750K).  Dashed contours indicate regions of negative intensity.  The grey scale is linear between -12 and 70 \jybm\ with dark areas indicating areas of emission and white areas indicating areas of absorption, and the image is centered on $\l=6, b=-1.1$.  \textit{Right:}  The same region from the 1.4 GHz Bonn single dish survey (Reich \etal~1990).  Note the nearby SNR W28 (G6.67$-$0.42, upper right) is seen in emission in both frames, but the \htwo\ regions M8 and W28A-2 are seen in emission at 1.4 GHz and seen in absorption at 74 MHz.  This is illustrative of the nonthermal nature of the SNR emission and the thermal absorption due to the \htwo\ regions.}
\label{fig:m8}
\end{figure}

After the calibration and imaging described in Section~\ref{sec:obs} were completed, the four fields were scrutinized for regions of negative intensity.  However, negative regions in an interferometer map can be image artifacts due to over-cleaning, incomplete \uv\ coverage, small errors in self-calibration, or other factors.  For this reason, an \textit{a priori} search was used to search for \htwo\ absorption features.  We used the catalog of Paladini et al.~(2003), which is a compilation of many separate \htwo\ region catalogs from higher frequency observations, for the identification.  Compact \htwo\ regions (smaller than 1.5\arcmin) were excluded 
as they would be undetectable due to beam dilution and the remaining \htwo\ regions were overlayed on the imaged regions.  Any known \htwo\ region that was coincident with an intensity $\leq -3\sigma$ was considered a detection.  The image intensity at the location of the \htwo\ region was then used to derive a sky brightness temperature\footnote{$I=\frac{2kT\Omega}{\lambda^{2}}$, where I is image intensity in $\frac{W}{m^2 Hz ster}$, $k$ is Boltzmann's constant, $\lambda$ is the wavelength, and $\Omega$ the solid angle of the beam.}.
Tables~2 through 5 detail the 92 absorption features that are detected.  No significant regions ($\ge 3$ contiguous beam areas) of negative intensity that did not coincide with a known \htwo\ region were found indicating that these absorption features are all astrophysical in origin\footnote{With 92 sources at $3\sigma$ confidence, one would expect less than one spurious detection.}  and not image artifacts.  Furthermore, the image median far from emission or absorption regions was measured to be near zero, confirming the assumption that the interferometer is not sensitive to the large scale emission.  A representative image of an \htwo\ absorption feature is shown in Figure~\ref{fig:m8}. 

\subsection{Nondetections and Detection Bias}\label{sec:bias}

Nearly 300 \htwo\ regions (of size $\ge 1.5\arcmin$) in the Paladini et al.~(2003) catalog and within the area imaged were not detected.  As \htwo\ absorption features are not detected via their own emission, but through the emission they are blocking, the sensitivity in detecting them depends on the temperature of the column behind them.  This is a particularly vexing detection bias as it not only varies as a function of Galactic longitude, Galactic latitude and distance, but the temperature of the background column is precisely the quantity in which we are interested, i.e. a model compensating for detection biases is the very model we are trying to create with the observations.  In this case and in cases where a large amount of the data in a given sample is contained in non-detections, the statistical techniques of survival (regression) analysis are of use (e.g. Isobe, Feigelson, \& Nelson 1986).  These techniques attempt to fit not only the measured data, but also extract information pertaining to the underlying distribution by assuming a functional form for the nondetection residuals, and fitting this well.  In order to perform this analysis, non detection upper limits must be measured.  The $3\sigma$ upper limits on the background temperature of the 134 \htwo\ regions in the field with known distances that were not detected are detailed in Appendix A.

\section{Deriving Cosmic Ray Emissivity Behind the HII Regions}\label{sec:derive}

From Section~\ref{sec:holes_theory}, the quantity directly measured from the images is $T_{obs}=T_e-T_{gb}$.  With knowledge of the electron temperatures of the \htwo\ regions, or lacking that adopting an assumed value (7000 K, see below), the integrated brightness temperature of the column behind the \htwo\ region can be derived.  Coupled with the distance to the \htwo\ region (when known), this gives many pencil beam integrated brightness temperatures with which to reconstruct the 3-dimensional CR electron emissivity distribution.  Note that this measurement is independent of assumptions of Galactic structure, a feature unique to this method.

Electron temperatures are either known from higher frequency studies (Paladini et al.~2004 and references therein) or may be estimated.  A value of $7000\pm2000$K was used as an estimate for the unknown electron temperatures based on the studies of Paladini et al.~(2004).  The trend for \htwo\ regions to have higher electron temperatures with increasing galactocentric radius (e.g.~Paladini et al.~2004 Figure 4) is not explicitly taken into account.  However the correlation is a weak one, and the narrow range of radii in this sample ($\sim$3-8 kpc) puts any systematic variation well within the errorbars. Thus for all \htwo\ regions detected in absorption an integrated brightness temperature of the column behind the \htwo\ region is derived; for nondetections, an upper limit is derived.  These values are detailed in the Tables 2 through 5 and in Appendix A.

\begin{figure}[htb]
\begin{center}
\includegraphics[scale=.9]{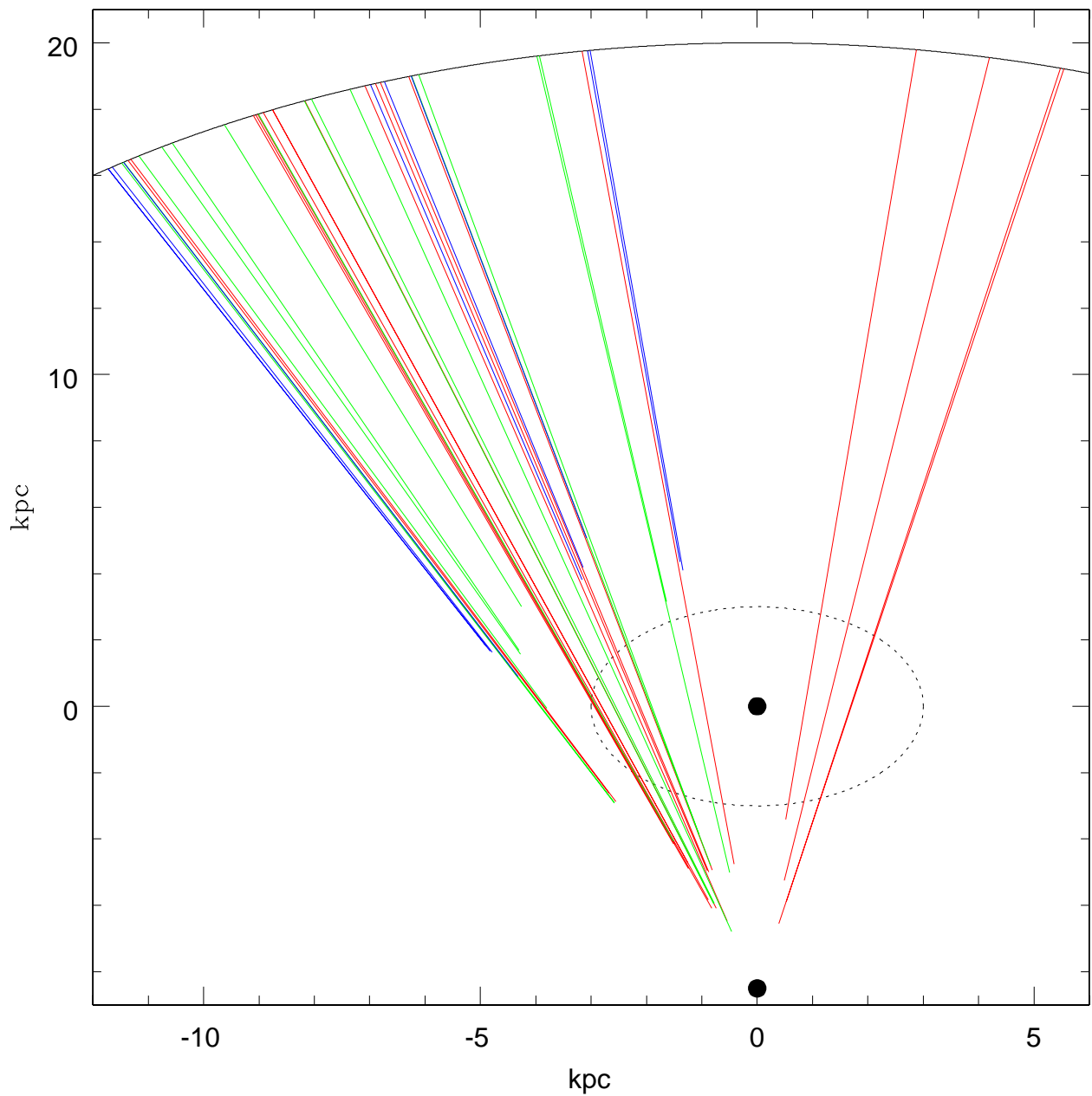}
\end{center}
\caption[Emissivity Lines of Sight]{Lines of sight with line-averaged emissivities measured through the \htwo\ absorption method.  The Galactic center is located at the origin and the earth is located at ($0,-8.5$) kpc (IAU 1985 standard).  The Galactic nonthermal emission is assumed to lie in a disk with radius 20 kpc (Ferri{\` e}re~2001 and references therein) as indicated by the solid black line.  Red lines represent lines of sight with emissivities $0.35< \ep <0.52$ \kpc, green lines represent $0.52< \ep < 0.72$, and blue lines represent $0.72 < \ep < 1.0$.  Note that the axes have different scales and the view is from Galactic North down onto the plane.  The dashed circle indicates the 3 kpc radius underdensity as modeled in Section~\ref{sec:model}.}
\label{fig:chords}
\end{figure}

For many of the \htwo\ regions in the survey area, kinematic distances have been determined  by Paladini et al. (2004 and references therein), who used absorption line data and a correlation between luminosity and linear diameter to resolve distance ambiguities when present.  Knowledge of the pathlength of the column behind the \htwo\ region along with the measured integrated brightness temperature allows for the average emissivity of the column behind the \htwo\ region to be derived ($\ep=T/D$ \kpc).  For the moment we model the Galactic CR disk as an axisymmetric cylinder having a radius of 20 kpc and a vertical height of 1 kpc above and below the plane (Ferri{\` e}re~2001 and references therein) and the earth is assumed to be 8.5 kpc from the Galactic center (IAU 1985 standard).  As all detected absorption features are $\le 1.3\arcdeg$ from the Galactic plane, all background lines of sight traverse the entire Galactic CR disk (i.e. no lines of sight exit the top or bottom of the disk) and therefore the disk may be treated 2-dimensionally.  The emissivities derived are listed in Tables 2 through 5.  Figure~\ref{fig:chords} shows graphically the lines of sight along which emissivities are measured.

As a check we compare our emissivity values with values derived from other methods.  Modeling of Galactic CR emissivity by Beuermann \etal\ (1985) based on the all-sky 408 MHz single dish maps of Haslam \etal\ (1982) derive a typical CR emissivity of $\sim10$~Kelvin~kpc$^{-1}$ at 408 MHz.  Planatia \etal\ (2003) examine the temperature spectral index of Galactic nonthermal emission through a comparison of the Haslam \etal\ (1982) 408 MHz map and the 1400 MHz map of Reich~(1982) and arrive at $T(\nu) \propto \nu^{-2.7}$.  With these values, the expected emissivity at 74 MHz may be estimated.  Scaling from 408 MHz, a value of $\sim$~1~Kelvin~pc$^{-1}$ is expected at 74 MHz.   The emissivity values measured in this work agree remarkably well with this estimate, with values ranging between $0.35 < \ep < 1.0$~\kpc.

\section{Galactic Cosmic Ray Emissivity Modeling}\label{sec:model}
  
The 42 emissivity path lengths (and 134 upper limits) measured in this work can form the starting point from which one may attempt to model the 3-dimensional Galactic CR emissivity.  The paucity of data and the uncertain detection biases involved make modeling extremely difficult.  Robust modeling based on these data may have to wait for the further development of low frequency radio telescope field, however in this Section we show one possible way in which such modeling might be undertaken.

The simplest possible model would be an axisymmetric disk of constant emissivity.  The first undertaking is to see if our data fit this model.  As discussed in Section~\ref{sec:bias}, a regression fit is required due to the multi-variate nature of the detection bias.  We use the linear regression with survival analysis using the EM Method\footnote{As implemented in IRAF Revision 2.12.2.  Data were also fit using the Buckley-James method with nearly identical results in all cases.}, which assumes a normal distribution of the non-detections around the fitted line.  This technique is reviewed extensively in Isobe, Feigelson, \& Nelson (1986).  The data were plotted as emissivity as a function of background column length\footnote{When a source was measured in more than one field, the mean of the two measured values was used}.  A fit with a zero slope would confirm the constant emissivity model, and the intercept of the fit gives the value of the emissivity.  The EM Method regression with survival analysis gives a fit of $\ep(D)=(-0.0109\pm0.0070)*D+0.5406\pm0.1654$, with $\ep$ in units of \kpc~ and D in kpc.  This fit is marginally ($1.5\sigma$) inconsistent with a zero slope, indicating the constant emissivity model may be incorrect.

Of note in Figure~\ref{fig:chords} is the result that the averaged
emissivities over path lengths originating on the far side of the Galactic
center have higher values on average than path lengths originating closer to the
observer.  As a simple departure from the above model, an underdensity in the CR electron emissivity in the inner Galaxy could explain the lower emissivity values seen over longer path lengths.

Utilizing the simple model of a cylinder 20 kpc in radius and 1 kpc thick, we calculate the mean emissivities one would expect to observe for a disk with uniform emissivity except for a central cylindrical region with zero emissivity.  We try radii of 1, 2, 3, 4, 5, and 6 kpc for the central hole.  EM Method regression with survival analysis fits were performed, with the results detailed in Table~\ref{tab:regression}.  For a central emissivity hole of radii 0, 1, and 2 kpc, the slope fit is not consistent with zero and is negative.  For values of 4, 5, and 6, the slope is inconsistent with zero and positive.  However, for the 3 kpc radius case, the fit is of the form $\ep(D)=(-0.0014\pm0.0073)*D+0.3646\pm0.1742$, a fit that is consistent with zero slope, indicating an average emissivity of $0.3646\pm0.1742$~\kpc\ for regions outside the central underdensity.  Figure~\ref{fig:emiss5} plots emissivity versus background path length for the observed data and the 3 kpc underdensity model.

Therefore, if a simple,
constant emissivity model is to be assumed, the best fit model entails a
region of zero emissivity approximately 3 kpc in radius and a constant
emissivity of ~$0.36\pm0.17$~\kpc\ outside this radius.  While we do not feel
that the data warrant a more complex model, our best fit model is clearly an
oversimplification.  In reality, for instance, the hole we have modeled is
unlikely to be devoid of emissivity and would therefore represent an
underdensity.

A literature search for works hypothesizing an underdensity in the Galactic
CR emissivities in the inner Galaxy has found no theoretical or
observational work indicating such a feature.  The 408 MHz three-dimensional
modeling of emissivity by Beuermann \etal\ (1985) has a slight underdensity at
a radius of about 3 kiloparsecs, but no central hole.  Modeling by Webber
\etal\ (1992) and Webber (1993a,b) based on CR nuclei observations and
propagation models suggest a weak (if any) radial dependence of CRs
and no central underdensity.  

Though little work has been done suggesting the
existence of a CR central underdensity, recent works (e.g.~Lorimer
2004, Case \& Bhattacharya 1998) suggest that there is a paucity of supernovae
remnants and pulsars in the inner several kpc.  Lorimer (2004) invokes the
current understanding of pulsar detection selection effects and applies these
to the recent and extremely successful Parkes Multibeam pulsar survey to
derive a pulsar radial density function which rises steeply from zero at the
Galactic center, peaks around 3.5 kpc and then falls exponentially (functional
form $\rho(R) = R^Ne^{-R/\sigma}$ Figure 4, Lorimer 2004).  Case \&
Bhattacharya (1998) use similar statistical methods as well as a
renormalization of the $\Sigma-D$ (Surface brightness-Distance) relationship
for shell SNRs to determine the density of SNRs as a function of galactocentric
radius.  A functional form similar to that of Lorimer (2004) is derived, with
maximum density at $\sim5$ kpc.  In short, new works on the distribution of astrophysical objects which are
derivatives of supernovae show a paucity in the few central
kpc, as opposed to a maximum or flat density (e.g. Narayan 1987).
If CR electrons are in fact closely associated with
SNRs, the central underdensity
indicated here would be consistent with an inner Galaxy underdensity of SNRs and
pulsars.  If the underdensity in CR emissivity
indicated by our simple model is confirmed by future work, we note that CR propagation processes are not capable of filling the hole.  This would be
an important consideration in future modeling of CR propagation.

We point out that none of the lines of sight discussed here pass within $\sim1$ kpc of the center of the Galaxy.  Therefore we are not hypothesizing a CR electron underdensity in the Galactic center.  High star formation rates in the central 50 pc (e.g.~Figer et al.~2004), and the diffuse nonthermal emission observed by LaRosa et al.~(2005) indicate that the central few hundred parsecs may possess emissivities higher than the values measured in the disk of the Galaxy.

This uniform emissivity model with a central underdensity is a simple example of the possible ways in which the 3-dimensional potential of \htwo\ absorption features could be utilized.  However with only 42 lines of sight available, more complex models (e.g. emissivity models which have radial or azimuthal dependence) are underdetermined and will have to await further observations.  

\section{Further Work}

It would be possible to increase the number of lines of sight available to
this technique through further VLA observations in the first Galactic quadrant
(the region extending into the fourth Galactic quadrant is inaccessible to the
VLA), or by determining the distances to more of the \htwo\ regions detected.
This second option is difficult in the vicinity of the Galactic center as the
radial velocities of \htwo\ regions are dominated by peculiar motions instead of the
overall Galactic rotation in this region.  It is also possible to calculate
the brightness temperature of the column between the observer and the \htwo\ region.
However, a measurement of the total emissivity along a nearby line-of-sight is
required.  This measurement would have to be sensitive to the total flux (the
\uv\ zero spacing) yet have a resolution and frequency similar to the
interferometer map.  The resolution would need to be similar (within a factor
of $\sim3$) because CR emissivity is sharply peaked along the Galactic
plane ($<1\arcdeg$ typical size; Beuermann et al.~1985).  Therefore large beam sizes result in beam dilution and an underestimation of
the brightness temperature.  The frequency needs to be similar ($50\lesssim \nu \lesssim
150$ MHz) as extrapolation over a large frequency range can cause large errors
and thermal emission and absorption change rapidly in this regime.  Existing
surveys, including the 408 MHz survey of Haslam \etal\ (1982),
the 1420 MHz survey of Reich (1982), and the 34.5 MHz survey of Dwarkanth \&
Udaya Shankar (1990), were considered, but do not have all the requirements to obtain a
reasonable estimate of the total intensity.  If such observations were to
exist, the brightness temperature (and emissivity when distance is known) of the column
in front of the \htwo\ region could be calculated, adding to the 3-dimensional
potential of this technique.

This technique may also hold some potential to break the degeneracy inherent in kinematic determinations of the distance to \htwo\ regions.  Because these 74 MHz \htwo\ absorption detections are sensitive to the integrated brightness temperature of the column \emph{behind} the absorbing medium (Section~\ref{sec:holes_theory}), the detections are biased towards closer \htwo\ regions.  Therefore, the \htwo\ regions
detected in this work are more likely "near" \htwo\ regions, i.e. the closer of the
two degenerate values.  This method was used by Brogan et al.~(2003) to determine the distance to the Sgr~D \htwo\ region.  Alternatively one could assume an emissivity along a line-of-sight, and attempt to match the observed depth of the \htwo\ absorption feature in order to get a rough estimate of distance.  However, background emissivity is not the only detection bias.  The range of spatial sensitivity will select for \htwo\ regions on intrinsically larger angular scales, i.e.~linearly larger and closer. An \htwo\ region that by chance had a strong CR electron source (i.e. SNR) behind it would have a larger $T_{bg}$ and therefore be easier to detect.  For these reasons, this method should only be considered an indicator, not proof that any given \htwo\ region has the closer of the distance degenerate values.

As a final note, we point out the CR electrons detected here through interaction with the Galactic magnetic field are also detected in the soft $\gamma$-ray ($\sim70$ MeV; Webber 1990) regime via bremsstrahlung with interstellar gas.  The radio measurement is essentially a line integration of CR density times magnetic field strength while the $\gamma$-ray measurement is a line integration of CR density times interstellar gas density.  Future radio measurements with emerging broad-band low
frequency arrays such as the Long Wavelength Array (LWA) will detect many
more \htwo\ absorption features and constrain the spectrum of the synchrotron emissivity.
Comparison of LWA and Gamma Ray Large Area Space Telescope (GLAST) data should then allow one to break the
degeneracy between magnetic field and cosmic ray electron distribution
inherent in radio observations alone, thereby providing direct information
on the strength and spatial distribution of the Galactic magnetic field.

\section{Conclusions}

It has long been known that it is possible to use the large optical depth of
\htwo\ regions at low radio frequencies to extract information concerning
the 3-dimensional distribution of CR emissivity without resorting to
making assumptions about Galactic structure. Though \htwo\ regions have been
observed in absorption at low frequencies before, the angular resolution and
sensitivity required to detect more than a few nearby sources have only recently become available using the 74 MHz system on the VLA.

Ninety two \htwo\ regions have been detected in absorption and 42 of these have
known distances, which are required to compute an emissivity.  The measured emissivities are consistent with values expected from higher frequency extrapolation, but are marginally inconsistent with a
constant cosmic ray emissivity model.  An emissivity model consisting of an axisymmetric disk 20 kpc in radius with emissivity $0.36\pm0.17$~\kpc\ between 3 and 20 kpc and zero inside 3 kpc is a better fit with the data.  Present data do not warrant more complex modeling (e.g. spiral arm structure, exponential disk, etc.).  The increased resolution and surface brightness sensitivity expected of the next generation of low frequency radio telescopes may be able to provide the increased number of lines-of-sight required for more complex modeling.

\acknowledgments

The authors thank Dr.~C.~Brogan of the University of Hawaii for allowing the use of previously collected data and assistance with data reduction, Dr.~T.N.~LaRosa for helpful comments, and the referee for very useful comments on selection effects and regression analysis.

Basic research in radio astronomy at the NRL is supported by the Office of Naval 
Research.  The National Radio Astronomy Observatory is a facility of the 
National Science Foundation operated under cooperative agreement by Associated 
Universities, Inc.


\setcounter{table}{0}%

\begin{deluxetable}{lcl}
\tablewidth{0pc}
\tablecaption{74 MHz Observational Summary}
\tablehead{\colhead{$(l,b)$}  &\colhead{Configuration} & \colhead{Date} }
\startdata
$(349.7$+$0.2)$    & DnC & 28 September 2001 \\   
                   &  C  & 28 August 2001    \\   
                   &  B  & 01 March 2001     \\   

$(0,0)$            &  D  & 31 May 1999       \\  
                   &  C  & 05 September 2001 \\  

$(11.2-0.3)$       & DnC & 28 September 2001 \\  
                   &  C  & 28 August 2001    \\  
                   &  B  & 01 March 2001     \\  

$(21.5-0.5)$       & DnC & 28 September 2001 \\  
                   &  C  & 31 August 2001    \\  
\enddata
\label{tab:74observe}

\end{deluxetable}

\cpap

\begin{deluxetable}{rrrllcccccc}
\label{tab:g21_holes}
\tabletypesize{\scriptsize}
\tablewidth{0pt}
\tablecaption{HII Regions in the Field of G21.5$-$0.5}
\tablecomments{Tables 2-5 detail the HII regions detected in absorption from Section~\ref{sec:holesdetect}.  Column (1) is the catalog number from Paladini \etal~(2003), Column (2) is Galactic longitude ($l$) in degrees, Column (3) is Galactic latitude ($b$) in degrees, Column (4) is Right Ascension (J2000), Column (5) is Declination (J2000), Column (6) is observed intensity in units \jybm, Column (7) is the sky brightness temperature derived from the measured intensity, Column (8) is the integrated cosmic ray electron brightness temperature of the column behind the HII region, Column (9) is the electron temperature of the HII region ($\times10^3$ Kelvin), Column (10) is the distance from the sun to the HII region (kiloparsecs, when known), and Column (11) is the emissivity (\kpc) of the column behind the HII region utilizing the assumptions of Section~\ref{sec:derive}.}
\tablecomments{Field centered on J2000 $18^{\mathrm{h}}33^{\mathrm{m}}36.4^{\mathrm{s}} -10\arcdeg 46\arcmin 38.8\arcsec$.  The resolution element is $7.33\arcmin \times 4.75\arcmin$, position angle 12\arcdeg (1\jybm=1790K).  RMS at the center of the field is 0.6~\jybm.  }

\tablehead{
\colhead{\# }      &\colhead{$l$}               &\colhead{$b$}                  & \colhead{RA}         &
\colhead{DEC}      & \colhead{I}                & \colhead{$T_{obs}$}           & \colhead{T$_{gb}$}   &
\colhead{T$_{e}$}  & \colhead{D}                &\colhead{$\ep$}                                \\

                   &                            &                        & \multicolumn{2}{c}{(J2000)}&
\colhead{(\jybm)}             &\multicolumn{3}{c}{($\times10^3$ Kelvin)} &\colhead{(kpc)}       &\colhead{(Kelvin~pc$^{-1}$)} \\
\colhead{(1)}&\colhead{(2)}&\colhead{(3)}&\colhead{(4)}&\colhead{(5)}&\colhead{(6)}&\colhead{(7)}&\colhead{(8)}
&\colhead{(9)}&\colhead{(10)}&\colhead{(11)}
}

\startdata

201	&$	15.1	$&$	-	0.7	$&	18	20	37.5	&$	-16	$	8	51	&$	-4.1	\pm	1.3	$&$	-7.3	\pm	2.3	$&$	13.2		\pm	2.5	$&$	5.9	\pm	1.0	$&$	2.1	$&$	0.51	\pm	0.10	$\\
204	&$	15.2	$&$	-	0.6	$&	18	20	27.2	&$	-16	$	0	43	&$	-4.4	\pm	1.3	$&$	-8.0	\pm	2.3	$&$	17.5		\pm	2.5	$&$	9.5	\pm	1.0	$&$	1.8	$&$	0.66	\pm	0.09	$\\
218	&$	16.9	$&$	+	0.8	$&	18	18	40.1	&$	-13	$	51	9	&$	-3.9	\pm	0.9	$&$	-6.9	\pm	1.7	$&$	13.0		\pm	1.9	$&$	6.1	\pm	1.0	$&$	2.7	$&$	0.52	\pm	0.08	$\\
219	&$	17.0	$&$	+	0.8	$&	18	18	51.8	&$	-13	$	45	52	&$	-3.9	\pm	0.9	$&$	-7.0	\pm	1.6	$&$	13.1		\pm	1.9	$&$	6.1	\pm	1.0	$&$	2.5	$&$	0.52	\pm	0.08	$\\
220	&$	17.0	$&$	+	0.9	$&	18	18	30.1	&$	-13	$	43	1	&$	-5.6	\pm	0.9	$&$	-10.0	\pm	1.6	$&$	18.1		\pm	1.9	$&$	8.1	\pm	1.0	$&$	2.7	$&$	0.72	\pm	0.08	$\\
224	&$	17.1	$&$	+	0.8	$&	18	19	3.5	&$	-13	$	40	34	&$	-5.4	\pm	0.9	$&$	-9.7	\pm	1.6	$&$	16.7	\tablenotemark{a}	\pm	2.6	$&$	7.0	\pm	2.0	$&$		$&$				$\\
234	&$	18.2	$&$	+	1.9	$&	18	17	13.1	&$	-12	$	11	13	&$	-2.4	\pm	0.8	$&$	-4.3	\pm	1.5	$&$	10.1		\pm	1.8	$&$	5.8	\pm	1.0	$&$		$&$				$\\
236	&$	18.3	$&$	-	0.3	$&	18	25	22.2	&$	-13	$	8	2	&$	-2.3	\pm	0.7	$&$	-4.2	\pm	1.3	$&$	9.5		\pm	1.6	$&$	5.3	\pm	1.0	$&$	4.0	$&$	0.40	\pm	0.07	$\\
237	&$	18.3	$&$	+	1.2	$&	18	19	56.2	&$	-12	$	25	47	&$	-3.2	\pm	0.8	$&$	-5.8	\pm	1.4	$&$	12.8	\tablenotemark{a}	\pm	2.4	$&$	7.0	\pm	2.0	$&$		$&$				$\\
238	&$	18.3	$&$	+	1.9	$&	18	17	24.7	&$	-12	$	5	56	&$	-3.1	\pm	0.8	$&$	-5.6	\pm	1.5	$&$	11.4		\pm	1.8	$&$	5.8	\pm	1.0	$&$	2.8	$&$	0.46	\pm	0.07	$\\
241	&$	18.5	$&$	+	1.9	$&	18	17	47.9	&$	-11	$	55	22	&$	-3.5	\pm	0.9	$&$	-6.3	\pm	1.7	$&$	13.3	\tablenotemark{a}	\pm	2.6	$&$	7.0	\pm	2.0	$&$		$&$				$\\
242	&$	18.5	$&$	+	2.0	$&	18	17	26.3	&$	-11	$	52	32	&$	-2.7	\pm	0.8	$&$	-4.8	\pm	1.5	$&$	11.8	\tablenotemark{a}	\pm	2.5	$&$	7.0	\pm	2.0	$&$		$&$				$\\
244	&$	18.6	$&$	+	1.9	$&	18	17	59.5	&$	-11	$	50	5	&$	-3.5	\pm	0.8	$&$	-6.3	\pm	1.4	$&$	13.3	\tablenotemark{a}	\pm	2.5	$&$	7.0	\pm	2.0	$&$		$&$				$\\
245	&$	18.7	$&$	+	2.0	$&	18	17	49.5	&$	-11	$	41	58	&$	-3.3	\pm	0.8	$&$	-6.0	\pm	1.5	$&$	13.0	\tablenotemark{a}	\pm	2.5	$&$	7.0	\pm	2.0	$&$	2.5	$&$	0.51	\pm	0.10	$\\
246	&$	18.8	$&$	+	1.8	$&	18	18	44.3	&$	-11	$	42	21	&$	-3.6	\pm	0.8	$&$	-6.5	\pm	1.4	$&$	13.5	\tablenotemark{a}	\pm	2.4	$&$	7.0	\pm	2.0	$&$		$&$				$\\
248	&$	18.9	$&$	-	0.5	$&	18	27	14.6	&$	-12	$	41	46	&$	-2.1	\pm	0.7	$&$	-3.8	\pm	1.2	$&$	9.7		\pm	1.6	$&$	5.9	\pm	1.0	$&$	4.6	$&$	0.42	\pm	0.07	$\\
249	&$	18.9	$&$	-	0.4	$&	18	26	52.8	&$	-12	$	38	59	&$	-5.1	\pm	0.7	$&$	-9.2	\pm	1.2	$&$	14.7		\pm	1.6	$&$	5.5	\pm	1.0	$&$	4.7	$&$	0.63	\pm	0.07	$\\
251	&$	19.0	$&$	-	0.4	$&	18	27	4.3	&$	-12	$	33	40	&$	-5.2	\pm	0.7	$&$	-9.3	\pm	1.2	$&$	16.3	\tablenotemark{a}	\pm	2.3	$&$	7.0	\pm	2.0	$&$		$&$				$\\
252	&$	19.0	$&$	-	0.3	$&	18	26	42.5	&$	-12	$	30	52	&$	-2.8	\pm	0.7	$&$	-5.1	\pm	1.2	$&$	10.3		\pm	1.6	$&$	5.2	\pm	1.0	$&$		$&$				$\\
253	&$	19.0	$&$	-	0.0	$&	18	25	37.3	&$	-12	$	22	29	&$	-1.9	\pm	0.7	$&$	-3.3	\pm	1.2	$&$	8.5		\pm	1.6	$&$	5.2	\pm	1.0	$&$	3.8	$&$	0.35	\pm	0.07	$\\
254	&$	19.1	$&$	-	0.3	$&	18	26	54.0	&$	-12	$	25	34	&$	-3.5	\pm	0.7	$&$	-6.3	\pm	1.2	$&$	10.4		\pm	1.6	$&$	4.1	\pm	1.0	$&$	4.6	$&$	0.45	\pm	0.07	$\\
266	&$	20.2	$&$	-	0.9	$&	18	31	9.7	&$	-11	$	43	49	&$	-1.7	\pm	0.6	$&$	-3.1	\pm	1.1	$&$	10.1	\tablenotemark{a}	\pm	2.3	$&$	7.0	\pm	2.0	$&$		$&$				$\\
267	&$	20.3	$&$	-	0.9	$&	18	31	21.1	&$	-11	$	38	30	&$	-2.0	\pm	0.6	$&$	-3.6	\pm	1.1	$&$	10.6	\tablenotemark{a}	\pm	2.3	$&$	7.0	\pm	2.0	$&$	12.3	$&$	0.69	\pm	0.15	$\\
268	&$	20.3	$&$	-	0.8	$&	18	30	59.3	&$	-11	$	35	43	&$	-2.4	\pm	0.6	$&$	-4.2	\pm	1.1	$&$	11.2	\tablenotemark{a}	\pm	2.3	$&$	7.0	\pm	2.0	$&$		$&$				$\\
286	&$	22.9	$&$	-	0.3	$&	18	34	4.0	&$	-9	$	3	30	&$	-2.9	\pm	0.6	$&$	-5.2	\pm	1.1	$&$	10.3		\pm	1.5	$&$	5.1	\pm	1.0	$&$	11.1	$&$	0.62	\pm	0.09	$\\
288	&$	23.0	$&$	-	0.4	$&	18	34	36.8	&$	-9	$	0	57	&$	-2.2	\pm	0.6	$&$	-4.0	\pm	1.1	$&$	11.8		\pm	1.5	$&$	7.8	\pm	1.0	$&$	10.9	$&$	0.71	\pm	0.09	$\\
309	&$	24.2	$&$	+	0.2	$&	18	34	41.7	&$	-7	$	40	27	&$	-2.3	\pm	0.7	$&$	-4.1	\pm	1.3	$&$	11.1	\tablenotemark{a}	\pm	2.4	$&$	7.0	\pm	2.0	$&$	9.3	$&$	0.61	\pm	0.13	$\\
311	&$	24.4	$&$	+	0.1	$&	18	35	25.5	&$	-7	$	32	34	&$	-2.9	\pm	0.7	$&$	-5.1	\pm	1.3	$&$	10.6		\pm	1.6	$&$	5.5	\pm	1.0	$&$	6.2	$&$	0.50	\pm	0.08	$\\
314	&$	24.5	$&$	+	0.2	$&	18	35	15.1	&$	-7	$	24	28	&$	-3.4	\pm	0.7	$&$	-6.1	\pm	1.3	$&$	10.8		\pm	1.7	$&$	4.7	\pm	1.0	$&$	6.5	$&$	0.51	\pm	0.08	$\\
316	&$	24.6	$&$	-	0.2	$&	18	36	52.2	&$	-7	$	30	11	&$	-4.0	\pm	0.7	$&$	-7.2	\pm	1.3	$&$	14.2	\tablenotemark{a}	\pm	2.4	$&$	7.0	\pm	2.0	$&$		$&$				$\\
319	&$	24.7	$&$	-	0.2	$&	18	37	3.4	&$	-7	$	24	51	&$	-4.7	\pm	0.7	$&$	-8.4	\pm	1.3	$&$	14.1		\pm	1.7	$&$	5.7	\pm	1.0	$&$	10.3	$&$	0.82	\pm	0.10	$\\
321	&$	24.7	$&$	-	0.1	$&	18	36	41.9	&$	-7	$	22	6	&$	-3.4	\pm	0.7	$&$	-6.2	\pm	1.3	$&$	13.2	\tablenotemark{a}	\pm	2.4	$&$	7.0	\pm	2.0	$&$	6.2	$&$	0.62	\pm	0.11	$\\
322	&$	24.8	$&$	+	0.1	$&	18	36	10.0	&$	-7	$	11	15	&$	-5.4	\pm	0.8	$&$	-9.8	\pm	1.4	$&$	14.7		\pm	1.7	$&$	5.0	\pm	1.0	$&$	6.1	$&$	0.69	\pm	0.08	$\\
325	&$	25.3	$&$	-	0.3	$&	18	38	31.5	&$	-6	$	55	38	&$	-4.2	\pm	0.8	$&$	-7.6	\pm	1.5	$&$	13.3		\pm	1.8	$&$	5.7	\pm	1.0	$&$	11.2	$&$	0.82	\pm	0.11	$\\
327	&$	25.4	$&$	-	0.3	$&	18	38	42.6	&$	-6	$	50	18	&$	-4.5	\pm	0.8	$&$	-8.0	\pm	1.5	$&$	16.1		\pm	1.8	$&$	8.1	\pm	1.0	$&$	11.2	$&$	1.00	\pm	0.11	$\\
328	&$	25.4	$&$	-	0.2	$&	18	38	21.1	&$	-6	$	47	33	&$	-4.6	\pm	0.8	$&$	-8.3	\pm	1.5	$&$	14.3		\pm	1.8	$&$	6.0	\pm	1.0	$&$	11.4	$&$	0.90	\pm	0.11	$\\

\enddata

\tablenotetext{a}{The electron temperature of this HII region was unavailable so a value of $7000\pm2000$K was used in the calculation of the background brightness temperature.}

\end{deluxetable}
\cpap

\begin{deluxetable}{rrrllcccccc}
\label{tab:g11_holes}
\tabletypesize{\scriptsize}
\tablewidth{0pt}
\tablecaption{HII Regions in the Field of G11.2$-$0.3}
\tablecomments{See notes in Table 2 for explanation of columns.} 
\tablecomments{Field centered on J2000 $18^{\mathrm{h}}11^{\mathrm{m}}30.600^{\mathrm{s}} -19\arcdeg 25\arcmin 16.000\arcsec$.  The resolution element is $6.67\arcmin \times 5.67\arcmin$, position angle $-10\arcdeg$ (1\jybm=1650K).  RMS at the center of the field is 0.7~\jybm.}

\tablehead{
\colhead{\# }      &\colhead{$l$}               &\colhead{$b$}                  & \colhead{RA}         &
\colhead{DEC}      & \colhead{I}                & \colhead{$T_{obs}$}           & \colhead{T$_{gb}$}   &
\colhead{T$_{e}$}  & \colhead{D}                &\colhead{$\ep$}                                \\

                   &                            &                        & \multicolumn{2}{c}{(J2000)}&
\colhead{(\jybm)}             &\multicolumn{3}{c}{($\times10^3$ Kelvin)} &\colhead{(kpc)}       &\colhead{(Kelvin~pc$^{-1}$)} \\
\colhead{(1)}&\colhead{(2)}&\colhead{(3)}&\colhead{(4)}&\colhead{(5)}&\colhead{(6)}&\colhead{(7)}&\colhead{(8)}
&\colhead{(9)}&\colhead{(10)}&\colhead{(11)}
}

\startdata

66	&$	6	$&$	-	1.3	$&	18	4	13	&$	-24	$	24	59	&$	-3.9	\pm	1.191	$&$	-6.4	\pm	2.0	$&$	13.4	\tablenotemark{a}	\pm	2.8	$&$	7.0	\pm	2.0	$&$		$&$				$\\
67	&$	6	$&$	-	1.2	$&	18	3	50	&$	-24	$	22	2	&$	-4.5	\pm	1.187	$&$	-7.4	\pm	2.0	$&$	14.4	\tablenotemark{a}	\pm	2.8	$&$	7.0	\pm	2.0	$&$		$&$				$\\
69	&$	6.1	$&$	-	0.6	$&	18	1	45.7	&$	-23	$	59	5	&$	-4.0	\pm	1.146	$&$	-6.7	\pm	1.9	$&$	13.7	\tablenotemark{a}	\pm	2.8	$&$	7.0	\pm	2.0	$&$	12.7	$&$	0.87	\pm	0.17	$\\
71	&$	6.2	$&$	-	1.2	$&	18	4	15.9	&$	-24	$	11	35	&$	-4.8	\pm	1.138	$&$	-8.0	\pm	1.9	$&$	15.0	\tablenotemark{a}	\pm	2.7	$&$	7.0	\pm	2.0	$&$		$&$				$\\
72	&$	6.2	$&$	-	0.6	$&	18	1	58.6	&$	-23	$	53	52	&$	-3.9	\pm	1.122	$&$	-6.4	\pm	1.9	$&$	13.4	\tablenotemark{a}	\pm	2.7	$&$	7.0	\pm	2.0	$&$	12.9	$&$	0.87	\pm	0.18	$\\
108	&$	8	$&$	-	0.2	$&	18	4	19	&$	-22	$	8	3	&$	-2.8	\pm	0.842	$&$	-4.6	\pm	1.4	$&$	11.6	\tablenotemark{a}	\pm	2.4	$&$	7.0	\pm	2.0	$&$	11.8	$&$	0.70	\pm	0.15	$\\
163	&$	12.7	$&$	-	0.2	$&	18	14	0.9	&$	-18	$	1	20	&$	-5.7	\pm	0.729	$&$	-9.4	\pm	1.2	$&$	13.9		\pm	1.6	$&$	4.5	\pm	1.0	$&$	4.8	$&$	0.59	\pm	0.07	$\\
165	&$	12.8	$&$	-	0.2	$&	18	14	13	&$	-17	$	56	4	&$	-5.6	\pm	0.733	$&$	-9.3	\pm	1.2	$&$	15.3		\pm	1.6	$&$	6.0	\pm	1.0	$&$	3.8	$&$	0.63	\pm	0.06	$\\
166	&$	12.8	$&$	+	0.4	$&	18	12	0.3	&$	-17	$	38	49	&$	-2.7	\pm	0.741	$&$	-4.5	\pm	1.2	$&$	11.5	\tablenotemark{a}	\pm	2.3	$&$	7.0	\pm	2.0	$&$	13.9	$&$	0.81	\pm	0.16	$\\
168	&$	12.9	$&$	-	0.2	$&	18	14	25	&$	-17	$	50	48	&$	-3.7	\pm	0.738	$&$	-6.0	\pm	1.2	$&$	12.1		\pm	1.6	$&$	6.1	\pm	1.0	$&$	3.7	$&$	0.49	\pm	0.06	$\\
177	&$	13.9	$&$	-	0.1	$&	18	16	2.9	&$	-16	$	55	12	&$	-2.9	\pm	0.800	$&$	-4.8	\pm	1.3	$&$	11.8	\tablenotemark{a}	\pm	2.4	$&$	7.0	\pm	2.0	$&$		$&$				$\\
178	&$	13.9	$&$	-	0	$&	18	15	40.8	&$	-16	$	52	21	&$	-3.5	\pm	0.800	$&$	-5.8	\pm	1.3	$&$	12.8	\tablenotemark{a}	\pm	2.4	$&$	7.0	\pm	2.0	$&$	13.1	$&$	0.85	\pm	0.16	$\\
181	&$	14	$&$	-	0.1	$&	18	16	14.8	&$	-16	$	49	56	&$	-4.3	\pm	0.808	$&$	-7.0	\pm	1.3	$&$	12.6		\pm	1.7	$&$	5.5	\pm	1.0	$&$	3.6	$&$	0.51	\pm	0.07	$\\
184	&$	14.2	$&$	-	0.3	$&	18	17	22.8	&$	-16	$	45	4	&$	-3.5	\pm	0.825	$&$	-5.8	\pm	1.4	$&$	12.8	\tablenotemark{a}	\pm	2.4	$&$	7.0	\pm	2.0	$&$		$&$				$\\
185	&$	14.2	$&$	-	0.2	$&	18	17	0.7	&$	-16	$	42	13	&$	-2.7	\pm	0.825	$&$	-4.4	\pm	1.4	$&$	11.4	\tablenotemark{a}	\pm	2.4	$&$	7.0	\pm	2.0	$&$	3.7	$&$	0.47	\pm	0.10	$\\
186	&$	14.2	$&$	-	0.1	$&	18	16	38.7	&$	-16	$	39	22	&$	-3.4	\pm	0.826	$&$	-5.6	\pm	1.4	$&$	12.6	\tablenotemark{a}	\pm	2.4	$&$	7.0	\pm	2.0	$&$		$&$				$\\
187	&$	14.3	$&$	-	0.2	$&	18	17	12.6	&$	-16	$	36	57	&$	-6.9	\pm	0.835	$&$	-11.4	\pm	1.4	$&$	18.4	\tablenotemark{a}	\pm	2.4	$&$	7.0	\pm	2.0	$&$		$&$				$\\
190	&$	14.4	$&$	-	0.6	$&	18	18	52.7	&$	-16	$	43	2	&$	-2.8	\pm	0.845	$&$	-4.6	\pm	1.4	$&$	11.6	\tablenotemark{a}	\pm	2.4	$&$	7.0	\pm	2.0	$&$		$&$				$\\
191	&$	14.4	$&$	-	0.1	$&	18	17	2.5	&$	-16	$	28	49	&$	-5.4	\pm	0.845	$&$	-9.0	\pm	1.4	$&$	16.0	\tablenotemark{a}	\pm	2.4	$&$	7.0	\pm	2.0	$&$	12.7	$&$	1.04	\pm	0.16	$\\
192	&$	14.5	$&$	-	0.6	$&	18	19	4.6	&$	-16	$	37	45	&$	-4.0	\pm	0.855	$&$	-6.6	\pm	1.4	$&$	13.6	\tablenotemark{a}	\pm	2.4	$&$	7.0	\pm	2.0	$&$		$&$				$\\
195	&$	14.6	$&$	+	0.1	$&	18	16	42.2	&$	-16	$	12	33	&$	-4.9	\pm	0.869	$&$	-8.1	\pm	1.4	$&$	13.4		\pm	1.7	$&$	5.3	\pm	1.0	$&$		$&$				$\\
218	&$	16.9	$&$	+	0.8	$&	18	18	40.1	&$	-13	$	51	9	&$	-5.6	\pm	1.356	$&$	-9.2	\pm	2.2	$&$	15.3		\pm	2.5	$&$	6.1	\pm	1.0	$&$	2.7	$&$	0.60	\pm	0.10	$\\

\enddata

\tablenotetext{a}{The electron temperature of this HII region was unavailable so a value of $7000\pm2000$K was used in the calculation of the background brightness temperature.}


\end{deluxetable}

\cpap

\begin{deluxetable}{rrrllcccccc}
\label{tab:gc_holes}
\tabletypesize{\scriptsize}
\tablewidth{0pt}
\tablecaption{HII Regions in the Field of the Galactic Center}
\tablecomments{See notes in Table 2 for explanation of columns.}
\tablecomments{Field centered on J2000 $17^{\mathrm{h}}45^{\mathrm{m}}40.0^{\mathrm{s}} -29\arcdeg 00\arcmin 27.9\arcsec$.  The resolution element is $17.41\arcmin \times 6.67\arcmin$, position angle $5\arcdeg$ (1\jybm=537K).  RMS at the center of the field is 0.8~\jybm.}

\tablehead{
\colhead{\# }      &\colhead{$l$}               &\colhead{$b$}                  & \colhead{RA}         &
\colhead{DEC}      & \colhead{I}                & \colhead{$T_{obs}$}           & \colhead{T$_{gb}$}   &
\colhead{T$_{e}$}  & \colhead{D}                &\colhead{$\ep$}                                \\

                   &                            &                        & \multicolumn{2}{c}{(J2000)}&
\colhead{(\jybm)}             &\multicolumn{3}{c}{($\times10^3$ Kelvin)} &\colhead{(kpc)}       &\colhead{(Kelvin~pc$^{-1}$)} \\
\colhead{(1)}&\colhead{(2)}&\colhead{(3)}&\colhead{(4)}&\colhead{(5)}&\colhead{(6)}&\colhead{(7)}&\colhead{(8)}
&\colhead{(9)}&\colhead{(10)}&\colhead{(11)}
}

\startdata

1	&$	0.1	$&$	+	0	$&	17	45	51.3	&$	-28	$	51	8	&$	-7.7	\pm	0.8	$&$	-4.1	\pm	0.4	$&$	11.1	\tablenotemark{a}	\pm	2.0	$&$	7.0	\pm	2.0	$&$		$&$				$\\
4	&$	0.3	$&$	-	0.5	$&	17	48	17	&$	-28	$	56	25	&$	-34.1	\pm	0.8	$&$	-18.3	\pm	0.4	$&$	25.3	\tablenotemark{a}	\pm	2.0	$&$	7.0	\pm	2.0	$&$		$&$				$\\
5	&$	0.4	$&$	-	0.8	$&	17	49	41.7	&$	-29	$	0	33	&$	-17.1	\pm	0.8	$&$	-9.2	\pm	0.4	$&$	16.2	\tablenotemark{a}	\pm	2.0	$&$	7.0	\pm	2.0	$&$		$&$				$\\
6	&$	0.4	$&$	-	0.5	$&	17	48	31.1	&$	-28	$	51	17	&$	-43.4	\pm	0.8	$&$	-23.3	\pm	0.4	$&$	30.8		\pm	1.1	$&$	7.5	\pm	1.0	$&$		$&$				$\\
7	&$	0.5	$&$	-	0.7	$&	17	49	32.2	&$	-28	$	52	19	&$	-21.0	\pm	0.8	$&$	-11.3	\pm	0.4	$&$	16.8		\pm	1.1	$&$	5.5	\pm	1.0	$&$		$&$				$\\
10	&$	0.6	$&$	-	0.9	$&	17	50	33.3	&$	-28	$	53	19	&$	-8.8	\pm	0.8	$&$	-4.7	\pm	0.4	$&$	11.7	\tablenotemark{a}	\pm	2.0	$&$	7.0	\pm	2.0	$&$		$&$				$\\
11	&$	0.6	$&$	-	0.6	$&	17	49	22.8	&$	-28	$	44	5	&$	-28.7	\pm	0.8	$&$	-15.4	\pm	0.4	$&$	22.4	\tablenotemark{a}	\pm	2.0	$&$	7.0	\pm	2.0	$&$		$&$				$\\
12	&$	0.6	$&$	-	0.5	$&	17	48	59.3	&$	-28	$	41	0	&$	-45.7	\pm	0.8	$&$	-24.6	\pm	0.4	$&$	38.1		\pm	1.1	$&$	13.5	\pm	1.0	$&$		$&$				$\\
13	&$	0.6	$&$	-	0.4	$&	17	48	35.9	&$	-28	$	37	54	&$	-45.4	\pm	0.8	$&$	-24.4	\pm	0.4	$&$	29.6		\pm	1.1	$&$	5.2	\pm	1.0	$&$		$&$				$\\
22	&$	2.3	$&$	+	1.4	$&	17	45	37.8	&$	-26	$	14	42	&$	-5.5	\pm	0.9	$&$	-3.0	\pm	0.5	$&$	10.0	\tablenotemark{a}	\pm	2.1	$&$	7.0	\pm	2.0	$&$		$&$				$\\
26	&$	2.4	$&$	+	1.4	$&	17	45	51.8	&$	-26	$	9	35	&$	-6.8	\pm	0.9	$&$	-3.6	\pm	0.5	$&$	10.6	\tablenotemark{a}	\pm	2.1	$&$	7.0	\pm	2.0	$&$		$&$				$\\
30	&$	3.3	$&$	+	0	$&	17	53	16.4	&$	-26	$	6	26	&$	-4.0	\pm	1.0	$&$	-2.1	\pm	0.5	$&$	9.1	\tablenotemark{a}	\pm	2.1	$&$	7.0	\pm	2.0	$&$		$&$				$\\
65	&$	5.9	$&$	-	0.4	$&	18	0	34	&$	-24	$	3	34	&$	-12.1	\pm	1.6	$&$	-6.5	\pm	0.9	$&$	13.2		\pm	1.3	$&$	6.7	\pm	1.0	$&$		$&$				$\\
66	&$	6	$&$	-	1.3	$&	18	4	13	&$	-24	$	24	59	&$	-12.3	\pm	1.7	$&$	-6.6	\pm	0.9	$&$	13.6	\tablenotemark{a}	\pm	2.2	$&$	7.0	\pm	2.0	$&$		$&$				$\\
67	&$	6	$&$	-	1.2	$&	18	3	50	&$	-24	$	22	2	&$	-19.0	\pm	1.7	$&$	-10.2	\pm	0.9	$&$	17.2	\tablenotemark{a}	\pm	2.2	$&$	7.0	\pm	2.0	$&$		$&$				$\\
69	&$	6.1	$&$	-	0.6	$&	18	1	45.7	&$	-23	$	59	5	&$	-13.3	\pm	1.7	$&$	-7.2	\pm	0.9	$&$	14.2	\tablenotemark{a}	\pm	2.2	$&$	7.0	\pm	2.0	$&$	12.7	$&$	0.90	\pm	0.14	$\\
71	&$	6.2	$&$	-	1.2	$&	18	4	15.9	&$	-24	$	11	35	&$	-20.1	\pm	1.8	$&$	-10.8	\pm	1.0	$&$	17.8	\tablenotemark{a}	\pm	2.2	$&$	7.0	\pm	2.0	$&$		$&$				$\\
72	&$	6.2	$&$	-	0.6	$&	18	1	58.6	&$	-23	$	53	52	&$	-14.9	\pm	1.7	$&$	-8.0	\pm	0.9	$&$	15.0	\tablenotemark{a}	\pm	2.2	$&$	7.0	\pm	2.0	$&$	12.9	$&$	0.97	\pm	0.14	$\\
75	&$	6.4	$&$	-	0.5	$&	18	2	1.7	&$	-23	$	40	28	&$	-6.8	\pm	1.8	$&$	-3.6	\pm	1.0	$&$	10.6	\tablenotemark{a}	\pm	2.2	$&$	7.0	\pm	2.0	$&$	3.8	$&$	0.43	\pm	0.09	$\\
78	&$	6.5	$&$	-	1.4	$&	18	5	40.3	&$	-24	$	1	46	&$	-8.2	\pm	2.0	$&$	-4.4	\pm	1.1	$&$	11.4	\tablenotemark{a}	\pm	2.3	$&$	7.0	\pm	2.0	$&$		$&$				$\\
109	&$	8.1	$&$	+	0.2	$&	18	3	1.6	&$	-21	$	51	1	&$	-15.8	\pm	3.3	$&$	-8.5	\pm	1.8	$&$	15.0		\pm	2.0	$&$	6.5	\pm	1.0	$&$	3.5	$&$	0.60	\pm	0.08	$\\
1401	&$	353.1	$&$	+	0.6	$&	17	25	43.2	&$	-34	$	25	19	&$	-11.2	\pm	2.1	$&$	-6.0	\pm	1.1	$&$	13.0	\tablenotemark{a}	\pm	2.3	$&$	7.0	\pm	2.0	$&$		$&$				$\\
1402	&$	353.1	$&$	+	0.7	$&	17	25	19.1	&$	-34	$	21	57	&$	-13.0	\pm	2.1	$&$	-7.0	\pm	1.1	$&$	14.0	\tablenotemark{a}	\pm	2.3	$&$	7.0	\pm	2.0	$&$		$&$				$\\
1403	&$	353.2	$&$	+	0.7	$&	17	25	35.4	&$	-34	$	16	59	&$	-20.5	\pm	2.0	$&$	-11.0	\pm	1.1	$&$	18.0	\tablenotemark{a}	\pm	2.3	$&$	7.0	\pm	2.0	$&$		$&$				$\\
1404	&$	353.2	$&$	+	0.9	$&	17	24	47.3	&$	-34	$	10	15	&$	-23.3	\pm	2.1	$&$	-12.5	\pm	1.1	$&$	19.5	\tablenotemark{a}	\pm	2.3	$&$	7.0	\pm	2.0	$&$		$&$				$\\
1407	&$	353.4	$&$	+	0.5	$&	17	26	56	&$	-34	$	13	44	&$	-17.5	\pm	1.9	$&$	-9.4	\pm	1.0	$&$	16.4	\tablenotemark{a}	\pm	2.2	$&$	7.0	\pm	2.0	$&$		$&$				$\\
1428	&$	357.5	$&$	-	1.4	$&	17	45	6.9	&$	-31	$	48	2	&$	-3.6	\pm	0.9	$&$	-2.0	\pm	0.5	$&$	9.0	\tablenotemark{a}	\pm	2.1	$&$	7.0	\pm	2.0	$&$		$&$				$\\
1429	&$	358	$&$	-	0.2	$&	17	41	33.8	&$	-30	$	44	42	&$	-5.1	\pm	0.9	$&$	-2.8	\pm	0.5	$&$	9.8	\tablenotemark{a}	\pm	2.1	$&$	7.0	\pm	2.0	$&$		$&$				$\\
1431	&$	358.6	$&$	-	0.1	$&	17	42	38.3	&$	-30	$	10	57	&$	-3.3	\pm	0.8	$&$	-1.8	\pm	0.4	$&$	8.8	\tablenotemark{a}	\pm	2.0	$&$	7.0	\pm	2.0	$&$		$&$				$\\
1440	&$	359.7	$&$	-	0.4	$&	17	46	28.1	&$	-29	$	24	6	&$	-3.1	\pm	0.8	$&$	-1.7	\pm	0.4	$&$	10.3		\pm	2.0	$&$	8.6	\pm	1.0	$&$		$&$				$\\
1441	&$	359.9	$&$	-	0.1	$&	17	45	46.2	&$	-29	$	4	30	&$	-14.1	\pm	0.8	$&$	-7.6	\pm	0.4	$&$	13.6		\pm	2.0	$&$	6.0	\pm	1.0	$&$		$&$				$\\
1442	&$	359.9	$&$	-	0	$&	17	45	22.8	&$	-29	$	1	22	&$	-20.4	\pm	0.8	$&$	-11.0	\pm	0.4	$&$	18.0	\tablenotemark{a}	\pm	2.0	$&$	7.0	\pm	2.0	$&$		$&$				$\\

\enddata

\tablenotetext{a}{The electron temperature of this HII region was unavailable so a value of $7000\pm2000$K was used in the calculation of the background brightness temperature.}


\end{deluxetable}

\cpap

\begin{deluxetable}{rrrllcccccc}
\label{tab:g349_holes}
\tabletypesize{\scriptsize}
\tablewidth{0pt}
\tablecaption{HII Regions in the Field of G349.7$+$0.2}
\tablecomments{See notes in Table 2 for explanation of columns.}
\tablecomments{Field centered on J2000 $17^{\mathrm{h}}18^{\mathrm{m}}02.700^{\mathrm{s}} -37\arcdeg 26\arcmin 13.0\arcsec$.  The resolution element is $8.33\arcmin \times 5.00\arcmin$, position angle $10\arcdeg$ (1\jybm=1490K).  RMS at the center of the field is 0.6~\jybm.}

\tablehead{
\colhead{\# }      &\colhead{$l$}               &\colhead{$b$}                  & \colhead{RA}         &
\colhead{DEC}      & \colhead{I}                & \colhead{$T_{obs}$}           & \colhead{T$_{gb}$}   &
\colhead{T$_{e}$}  & \colhead{D}                &\colhead{$\ep$}                                \\

                   &                            &                        & \multicolumn{2}{c}{(J2000)}&
\colhead{(\jybm)}             &\multicolumn{3}{c}{($\times10^3$ Kelvin)} &\colhead{(kpc)}       &\colhead{(Kelvin~pc$^{-1}$)} \\
\colhead{(1)}&\colhead{(2)}&\colhead{(3)}&\colhead{(4)}&\colhead{(5)}&\colhead{(6)}&\colhead{(7)}&\colhead{(8)}
&\colhead{(9)}&\colhead{(10)}&\colhead{(11)}
}

\startdata

1353	&$	348.6	$&$	-	0.6	$&	17	17	53.5	&$	-38	$	48	24	&$	-5.5	\pm	0.6	$&$	-8.2	\pm	0.9	$&$	13.0		\pm	1.4	$&$	4.8	\pm	1.0	$&$	2.7	$&$	0.51	\pm	0.05	$\\
1354	&$	348.7	$&$	-	1	$&	17	19	52.2	&$	-38	$	57	17	&$	-2.2	\pm	0.6	$&$	-3.3	\pm	0.9	$&$	9.5		\pm	1.4	$&$	6.2	\pm	1.0	$&$	2.0	$&$	0.36	\pm	0.05	$\\
1375	&$	351	$&$	+	0.7	$&	17	19	29.2	&$	-36	$	5	44	&$	-4.6	\pm	0.6	$&$	-6.8	\pm	0.9	$&$	13.8	\tablenotemark{a}	\pm	2.2	$&$	7	\pm	2.0	$&$		$&$				$\\
1376	&$	351.2	$&$	+	0.5	$&	17	20	51.9	&$	-36	$	2	45	&$	-4.4	\pm	0.6	$&$	-6.7	\pm	0.9	$&$	13.7	\tablenotemark{a}	\pm	2.2	$&$	7	\pm	2.0	$&$		$&$				$\\
1379	&$	351.4	$&$	+	0.7	$&	17	20	37.1	&$	-35	$	46	3	&$	-4.2	\pm	0.6	$&$	-6.2	\pm	0.9	$&$	12.3		\pm	1.4	$&$	6.1	\pm	1.0	$&$		$&$				$\\
1380	&$	351.5	$&$	-	0.5	$&	17	25	47.7	&$	-36	$	21	49	&$	-2.1	\pm	0.6	$&$	-3.2	\pm	1.0	$&$	8.9		\pm	1.4	$&$	5.7	\pm	1.0	$&$	3.3	$&$	0.36	\pm	0.06	$\\
1411	&$	354.2	$&$	-	0.1	$&	17	31	28.9	&$	-33	$	53	39	&$	-3.3	\pm	0.9	$&$	-4.9	\pm	1.3	$&$	10.2		\pm	1.6	$&$	5.3	\pm	1.0	$&$	5.1	$&$	0.44	\pm	0.07	$\\
1412	&$	354.2	$&$	-	0	$&	17	31	4.8	&$	-33	$	50	21	&$	-2.9	\pm	0.9	$&$	-4.3	\pm	1.3	$&$	9.6		\pm	1.6	$&$	5.3	\pm	1.0	$&$		$&$				$\\

\enddata

\tablenotetext{a}{The electron temperature of this HII region was unavailable so a value of $7000\pm2000$K was used in the calculation of the background brightness temperature.}

\end{deluxetable}

\cpap

\begin{deluxetable}{crr}
\tablewidth{0pc}
\tablecaption{EM Method Linear Regression With Survival Analysis Fits}
\tablecomments{EM Method regression with survival analysis fits for the model fitting in Section~\ref{sec:model}.  Column (1) is the size in kiloparsecs of the hypothesized central region devoid of emissivity.  Columns (2) and (3) are the slope and intercept of the fit.}
\tablehead{\colhead{R}      &\colhead{Slope} & \colhead{Intercept} \\
           \colhead{(kpc)}  &\colhead{(K)}     & \colhead{($10^{-3}$~Kelvin~pc$^{-2}$)}  \\
           \colhead{(1)}  &\colhead{(2)}     & \colhead{(3)}
}
\startdata

0	&$-0.0109\pm0.0070$	&$0.5406\pm0.1654$	\\

1	&$-0.0108\pm0.0070$	&$0.5401\pm0.1652$	\\

2	&$-0.0091\pm0.0069$	&$0.5129\pm0.1639$	\\

3	&$-0.0014\pm0.0073$	&$0.3646\pm0.1742$	\\

4	&$0.0078\pm0.0082$	&$0.1903\pm0.1946$	\\

5	&$0.0120\pm0.0092$	&$0.1322\pm0.2188$	\\

6	&$0.0144\pm0.0101$	&$0.1145\pm0.2394$	\\

\enddata
\label{tab:regression}

\end{deluxetable}

\cpap


\cpap

\cpap

\cpap

\begin{figure}[htb]
\begin{center}
\includegraphics[scale=.9]{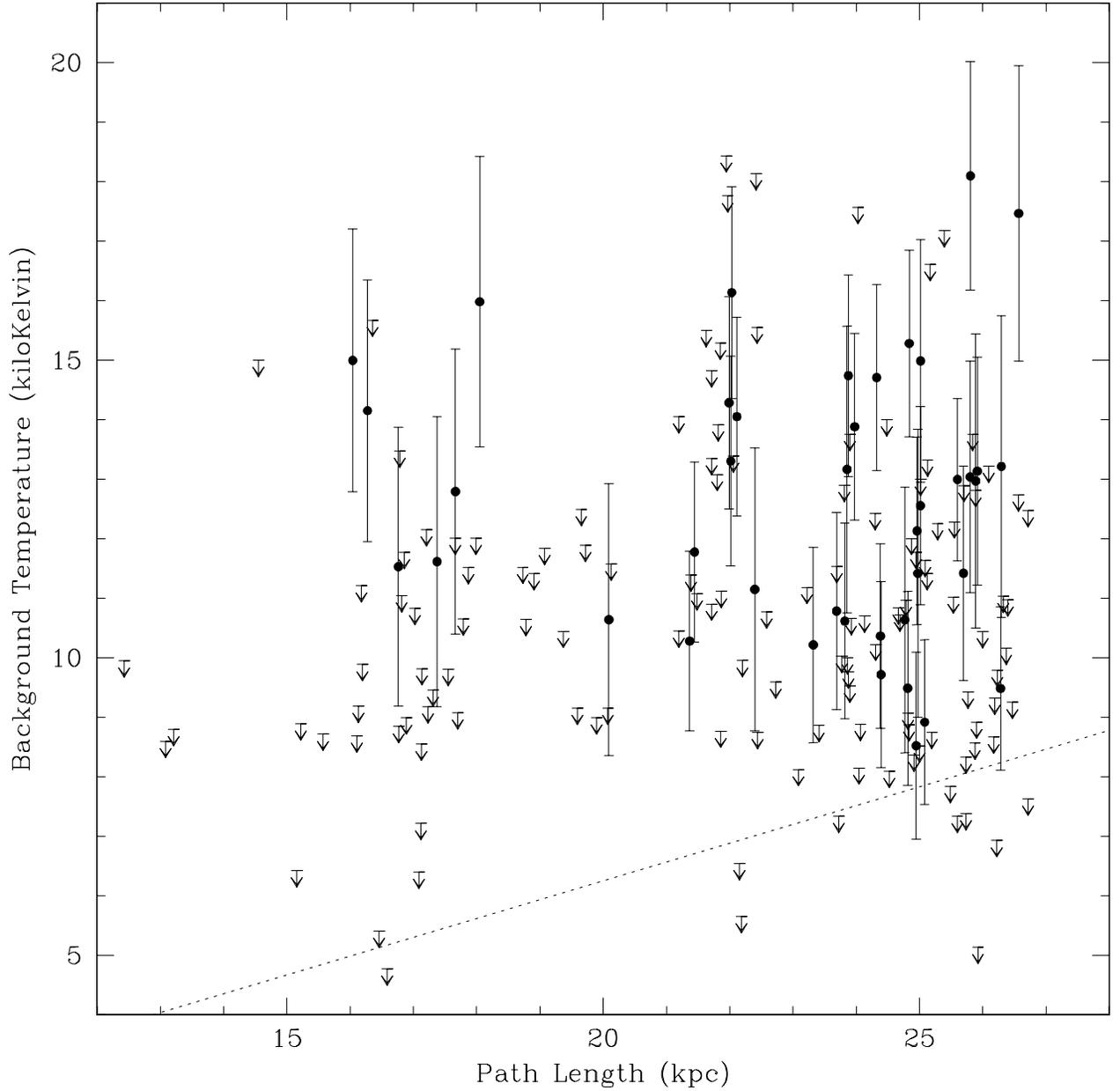}
\end{center}
\caption[]{Background path length versus background temperature plotted with Galactic CR disk of radius 20 kpc.  Circles are detections, downward arrows are $3\sigma$ upper limits, and the dotted line is the EM Method linear regression with survival analysis fit to the data.}
\label{fig:t_d}
\end{figure}
\cpap

\begin{figure}[htb]
\begin{center}
\includegraphics[scale=.9]{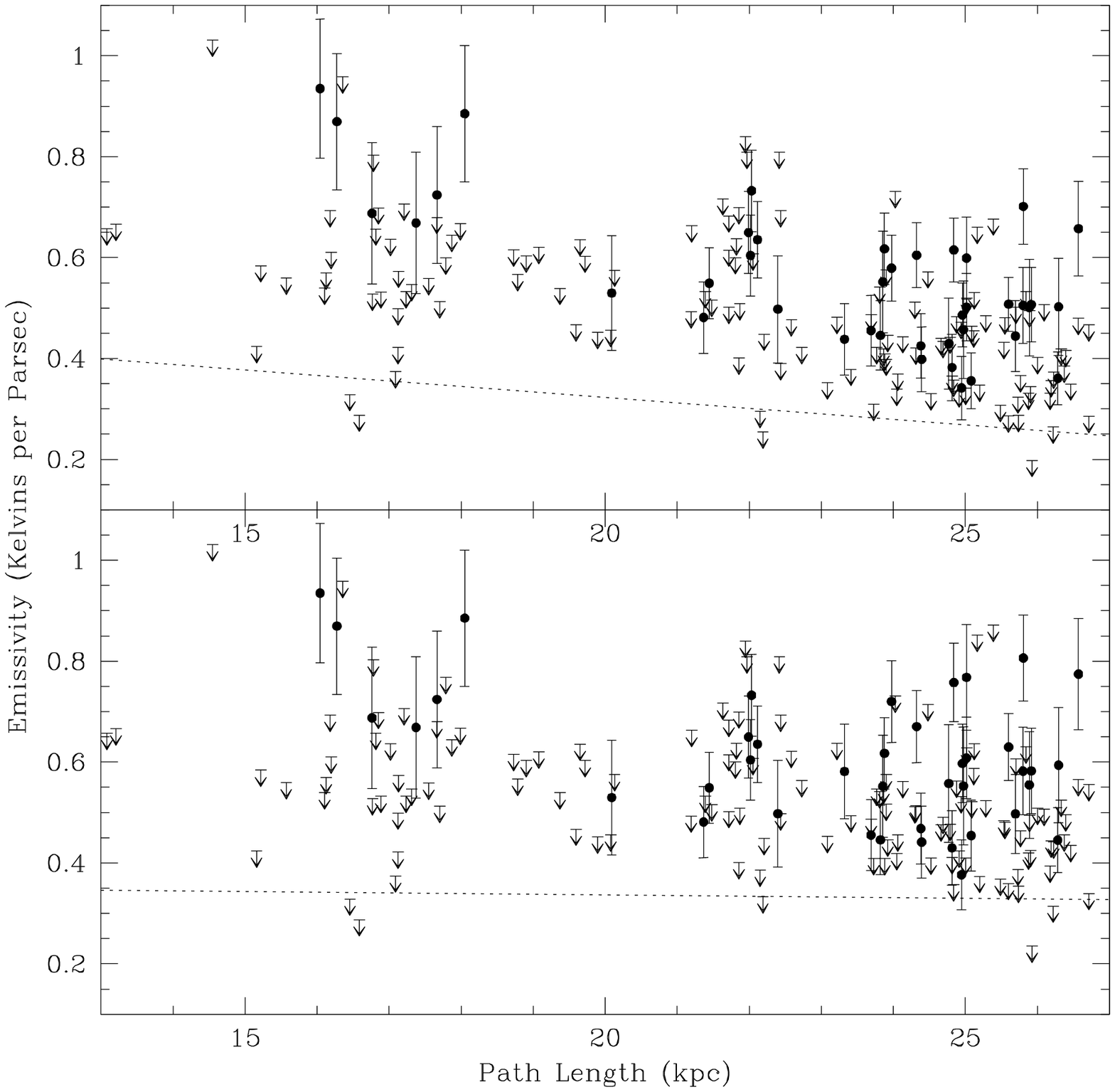}
\end{center}
\caption[Emissivities Derived from the inner Galaxy Emissivity Underdensity
Model]{Emissivities derived from the inner Galaxy emissivity underdensity model.
Values on the ordinate are integrated emissivities in \kpc.  Values on the
axis are path length of the emissivity measurement in kiloparsecs.  Detections are indicated with circles, $3\sigma$ upper limits with downward arrows, and the dotted line is the regression fit.  \textit{Top:}  No central underdensity.  The regression fit is marginally ($1.5\sigma$) inconsistent with a uniform emissivity model.  \textit{Bottom:}  3 kiloparsec central underdensity model.  The regression fit is consistent with constant emissivity.}
\label{fig:emiss5}

\end{figure}
\cpap

\appendix

\section{HII Region Non-Detections}

\begin{deluxetable}{rrrrrr}
\label{tab:nondetect}
\tabletypesize{\scriptsize}
\tablewidth{0pt}
\tablecaption{Appendix A:  HII Region Non-Detections}
\tablecomments{Column (1) is the catalog number from Paladini \etal~(2003), Column (2) is Galactic longitude ($l$) in degrees, Column (3) is Galactic latitude ($b$) in degrees, Column (4) is geocentric distance (kpc), Column (5) is electron temperature ($\times10^3$ Kelvin, when available), and Column (6) is the $3\sigma$ nondetection upper limit of the integrated cosmic ray electron brightness temperature of the column behind the HII region ($\times10^3$ Kelvin, 7000 Kelvin used for electron temperature when not available).}

\tablehead{
\colhead{\# }      &\colhead{$l$}               &\colhead{$b$}    &\colhead{D}     &\colhead{T$_{e}$}     & \colhead{$T_{gb}$}   \\

                   &                            &                 &\colhead{(kpc)} &   \multicolumn{2}{c}{($\times10^3$ Kelvin)} \\

\colhead{(1)}&\colhead{(2)}&\colhead{(3)}&\colhead{(4)}&\colhead{(5)}&\colhead{(6)}
}

\startdata

43	&	4.6	&$	-	0.1	$&	12.3	&	3.6	&	5.4	\\
48	&	5	&$	+	0.3	$&	13.6	&	\nodata	&	8.9	\\
60	&	5.5	&$	-	0.2	$&	4.8	&	6.4	&	7.3	\\
80	&	6.5	&$	+	0.1	$&	3.1	&	7.5	&	17.2	\\
81	&	6.6	&$	-	0.3	$&	3.4	&	5.8	&	16.6	\\
82	&	6.6	&$	-	0.1	$&	3.0	&	6.7	&	22.5	\\
86	&	6.7	&$	-	0.2	$&	3.7	&	8.1	&	36.5	\\
93	&	7	&$	-	0.2	$&	3.0	&	7.4	&	28.0	\\
100	&	7.4	&$	+	0.7	$&	13.1	&	4.7	&	8.7	\\
112	&	8.3	&$	-	0.1	$&	11.6	&	\nodata	&	9.8	\\
114	&	8.4	&$	-	0.3	$&	4.7	&	7.3	&	9.5	\\
115	&	8.5	&$	-	0.3	$&	12.1	&	\nodata	&	12.2	\\
117	&	8.7	&$	-	0.4	$&	4.9	&	7.8	&	10.0	\\
119	&	8.9	&$	-	0.3	$&	12.6	&	\nodata	&	8.8	\\
123	&	9.7	&$	-	0.8	$&	12.7	&	\nodata	&	9.0	\\
128	&	10	&$	-	0.8	$&	12.5	&	3.7	&	6.4	\\
129	&	10.1	&$	-	0.4	$&	14.5	&	6.0	&	8.7	\\
131	&	10.2	&$	-	0.4	$&	2.9	&	5.6	&	7.3	\\
132	&	10.2	&$	-	0.3	$&	2.3	&	5.4	&	9.3	\\
133	&	10.3	&$	-	0.3	$&	12.7	&	\nodata	&	10.8	\\
134	&	10.3	&$	-	0.2	$&	2.0	&	6.7	&	9.2	\\
135	&	10.3	&$	-	0.1	$&	11.5	&	\nodata	&	12.0	\\
137	&	10.5	&$		0.0	$&	5.7	&	6.0	&	8.1	\\
140	&	10.7	&$		0.0	$&	12.7	&	\nodata	&	8.5	\\
142	&	10.9	&$	+	0.1	$&	14.0	&	\nodata	&	9.2	\\
143	&	11	&$		0.0	$&	14.0	&	\nodata	&	11.2	\\
145	&	11.2	&$	-	1.1	$&	13.5	&	3.8	&	4.8	\\
150	&	11.7	&$	-	1.7	$&	2.6	&	\nodata	&	8.9	\\
154	&	11.9	&$		0.0	$&	4.2	&	8.6	&	14.0	\\
155	&	11.9	&$	+	0.7	$&	13.5	&	7.9	&	11.0	\\
158	&	12	&$	-	0.2	$&	12.1	&	\nodata	&	11.5	\\
159	&	12.4	&$	-	1.1	$&	4.1	&	\nodata	&	8.1	\\
160	&	12.4	&$		0.0	$&	21.8	&	\nodata	&	9.9	\\
167	&	12.9	&$	-	0.3	$&	13.3	&	4.7	&	9.2	\\
169	&	13.2	&$	-	0.1	$&	4.7	&	6.4	&	8.9	\\
170	&	13.2	&$		0.0	$&	4.7	&	4.9	&	8.1	\\
173	&	13.4	&$	+	0.1	$&	14.2	&	4.7	&	13.5	\\
174	&	13.5	&$	-	0.2	$&	14.9	&	\nodata	&	15.7	\\
176	&	13.8	&$	-	0.8	$&	2.8	&	8.6	&	12.9	\\
180	&	13.9	&$	+	0.3	$&	4.4	&	6.2	&	10.2	\\
182	&	14.1	&$	-	0.5	$&	14.0	&	\nodata	&	9.8	\\
186	&	14.2	&$	-	0.1	$&	3.6	&	\nodata	&	13.0	\\
188	&	14.3	&$	+	0.1	$&	13.3	&	\nodata	&	12.0	\\
189	&	14.4	&$	-	0.7	$&	1.6	&	10.4	&	12.5	\\
194	&	14.6	&$		0.0	$&	3.5	&	4.7	&	13.3	\\
195	&	14.6	&$	+	0.1	$&	3.6	&	4.6	&	8.5	\\
197	&	14.7	&$	-	0.5	$&	13.4	&	\nodata	&	9.1	\\
199	&	15	&$	-	0.7	$&	2.0	&	6.9	&	10.2	\\
200	&	15.1	&$	-	0.9	$&	1.8	&	8.7	&	12.7	\\
203	&	15.2	&$	-	0.8	$&	2.6	&	10.0	&	13.8	\\
209	&	16.3	&$	-	0.2	$&	12.2	&	\nodata	&	11.4	\\
210	&	16.4	&$	-	0.5	$&	12.6	&	\nodata	&	11.5	\\
211	&	16.4	&$	-	0.2	$&	12.5	&	6.5	&	10.6	\\
212	&	16.6	&$	-	0.3	$&	3.8	&	\nodata	&	12.0	\\
230	&	18.1	&$	-	0.3	$&	4.1	&	6.3	&	10.7	\\
231	&	18.2	&$	-	0.4	$&	3.6	&	6.2	&	11.6	\\
232	&	18.2	&$	-	0.3	$&	3.8	&	7.6	&	11.8	\\
233	&	18.2	&$	-	0.2	$&	12.4	&	\nodata	&	10.4	\\
235	&	18.3	&$	-	0.4	$&	3.0	&	6.7	&	11.0	\\
236	&	18.3	&$	-	0.3	$&	4.0	&	5.3	&	9.1	\\
240	&	18.5	&$		0.0	$&	11.9	&	\nodata	&	11.9	\\
243	&	18.6	&$	-	0.3	$&	11.3	&	\nodata	&	9.2	\\
250	&	19	&$	-	0.6	$&	11.4	&	\nodata	&	11.6	\\
257	&	19.5	&$	+	0.1	$&	14.0	&	7.7	&	11.8	\\
258	&	19.6	&$	-	0.2	$&	3.4	&	7.9	&	12.3	\\
259	&	19.6	&$	-	0.1	$&	4.2	&	7.3	&	10.8	\\
260	&	19.7	&$	-	0.2	$&	3.5	&	5.3	&	8.7	\\
261	&	19.7	&$	-	0.1	$&	4.1	&	7.9	&	11.0	\\
265	&	20.1	&$	-	0.1	$&	12.6	&	7.1	&	9.0	\\
271	&	20.5	&$	+	0.2	$&	13.7	&	\nodata	&	9.2	\\
272	&	20.7	&$	-	0.1	$&	4.0	&	6.3	&	8.9	\\
274	&	21	&$	+	0.1	$&	14.1	&	8.2	&	12.5	\\
280	&	21.9	&$	-	0.4	$&	10.8	&	\nodata	&	10.4	\\
281	&	21.9	&$		0.0	$&	2.2	&	\nodata	&	13.2	\\
283	&	22.4	&$	+	0.1	$&	10.5	&	\nodata	&	11.1	\\
284	&	22.8	&$	-	0.5	$&	10.9	&	6.1	&	11.4	\\
285	&	22.8	&$	-	0.3	$&	5.0	&	6.4	&	12.4	\\
291	&	23.1	&$	-	0.2	$&	10.3	&	\nodata	&	13.3	\\
293	&	23.1	&$	+	0.6	$&	2.5	&	\nodata	&	12.8	\\
295	&	23.4	&$	-	0.2	$&	5.8	&	6.3	&	13.8	\\
296	&	23.5	&$		0.0	$&	10.2	&	4.3	&	8.8	\\
297	&	23.6	&$	-	0.4	$&	10.3	&	\nodata	&	11.1	\\
299	&	23.7	&$	-	0.2	$&	10.7	&	\nodata	&	10.9	\\
300	&	23.7	&$	+	0.2	$&	5.9	&	5.9	&	9.8	\\
301	&	23.8	&$	+	0.2	$&	9.5	&	5.8	&	10.0	\\
302	&	23.9	&$	-	0.1	$&	10.9	&	6.3	&	14.8	\\
303	&	23.9	&$	+	0.1	$&	11.1	&	\nodata	&	15.5	\\
305	&	24	&$	+	0.2	$&	10.7	&	5.9	&	13.9	\\
306	&	24.1	&$	-	0.1	$&	10.3	&	\nodata	&	17.8	\\
307	&	24.1	&$	+	0.1	$&	9.2	&	\nodata	&	18.1	\\
308	&	24.2	&$	-	0.1	$&	10.4	&	7.3	&	18.4	\\
310	&	24.3	&$	-	0.2	$&	10.8	&	\nodata	&	15.3	\\
313	&	24.5	&$		0.0	$&	9.3	&	\nodata	&	15.6	\\
314	&	24.5	&$	+	0.2	$&	6.5	&	4.7	&	11.5	\\
315	&	24.5	&$	+	0.5	$&	5.8	&	6.1	&	17.6	\\
319	&	24.7	&$	-	0.2	$&	6.3	&	5.8	&	12.9	\\
324	&	25.2	&$	+	0.1	$&	11.9	&	\nodata	&	13.1	\\
325	&	25.3	&$	-	0.3	$&	3.0	&	7.5	&	12.3	\\
329	&	25.4	&$		0.0	$&	16.7	&	7.9	&	14.1	\\
331	&	25.7	&$		0.0	$&	11.5	&	7.7	&	13.4	\\
332	&	25.8	&$	+	0.2	$&	6.3	&	5.0	&	10.7	\\
1326	&	345.2	&$	-	0.7	$&	2.2	&	4.0	&	5.1	\\
1328	&	345.3	&$	+	1.5	$&	1.9	&	7.2	&	9.8	\\
1328	&	345.3	&$	+	1.5	$&	1.9	&	6.0	&	8.7	\\
1330	&	345.4	&$	-	0.9	$&	2.6	&	6.2	&	7.8	\\
1331	&	345.4	&$	+	1.4	$&	1.9	&	4.7	&	6.9	\\
1332	&	345.5	&$	+	0.2	$&	1.7	&	8.4	&	11.0	\\
1333	&	345.5	&$	+	0.3	$&	2.4	&	7.2	&	8.3	\\
1335	&	345.6	&$		0.0	$&	1.4	&	6.5	&	7.6	\\
1336	&	345.8	&$		0.0	$&	15.1	&	7.4	&	8.6	\\
1337	&	346.1	&$		0.0	$&	14.9	&	5.7	&	8.8	\\
1338	&	346.2	&$	-	0.1	$&	10.4	&	9.0	&	10.7	\\
1342	&	347.4	&$	+	0.3	$&	6.1	&	4.6	&	6.5	\\
1343	&	347.6	&$	+	0.2	$&	6.0	&	3.9	&	5.7	\\
1347	&	347.9	&$		0.0	$&	13.1	&	4.7	&	6.4	\\
1350	&	348	&$	-	0.4	$&	5.8	&	6.2	&	8.7	\\
1351	&	348.2	&$	-	1.0	$&	2.5	&	5.0	&	7.4	\\
1356	&	349.1	&$		0.0	$&	4.2	&	\nodata	&	10.7	\\
1357	&	349.1	&$	+	0.1	$&	11.0	&	7.4	&	9.5	\\
1361	&	349.8	&$	-	0.5	$&	3.4	&	6.3	&	8.4	\\
1364	&	350.1	&$	+	0.1	$&	5.7	&	8.7	&	10.8	\\
1366	&	350.5	&$	+	1.0	$&	2.0	&	9.0	&	11.0	\\
1373	&	351	&$	-	0.6	$&	13.8	&	6.1	&	15.0	\\
1381	&	351.6	&$	-	1.3	$&	2.6	&	7.0	&	9.4	\\
1382	&	351.6	&$	-	1.2	$&	2.4	&	6.0	&	10.4	\\
1384	&	351.6	&$	+	0.2	$&	5.0	&	5.8	&	8.9	\\
1385	&	351.7	&$	-	1.2	$&	2.5	&	5.7	&	8.6	\\
1394	&	352.7	&$	+	0.1	$&	11.5	&	10.5	&	11.8	\\
1397	&	352.9	&$	-	0.2	$&	11.3	&	5.4	&	7.2	\\
1405	&	353.4	&$	-	0.4	$&	3.3	&	7.2	&	11.4	\\
1406	&	353.4	&$	-	0.1	$&	5.2	&	\nodata	&	11.2	\\
1408	&	353.5	&$		0.0	$&	5.7	&	3.3	&	9.6	\\
1410	&	353.6	&$		0.0	$&	4.6	&	7.0	&	10.0	\\
1417	&	354.7	&$	+	0.5	$&	12.3	&	5.3	&	9.9	\\

\enddata

\end{deluxetable}

\end{document}